\documentclass[aps,prd,onecolumn,groupedaddress,showpacs,nofootinbib,amssymb]{revtex4}
\usepackage[dvips]{graphicx}
\usepackage{amssymb}
\usepackage{amsmath}
\usepackage{graphicx,,color}
\usepackage{amsfonts}
\usepackage{bm}
\usepackage{cancel}
\usepackage{comment}
\usepackage{listings}

\newcommand\be{\begin{equation}}
\newcommand\ee{\end{equation}}

\allowdisplaybreaks[4]

\begin{document}

\title{Unifying Inflation with Early and Late Dark Energy Epochs in Axion $F(R)$ Gravity}
\author{V.K.
Oikonomou,$^{1,2}$\,\thanks{v.k.oikonomou1979@gmail.com}}
\affiliation{$^{1)}$ Department of Physics, Aristotle University
of Thessaloniki, Thessaloniki 54124,
Greece\\
$^{2)}$ Laboratory for Theoretical Cosmology, Tomsk State
University of Control Systems and Radioelectronics, 634050 Tomsk,
Russia (TUSUR)}

\tolerance=5000

\begin{abstract}
We provide a theoretical model of $F(R)$ gravity in which it is
possible to describe in a unified way inflation, an early and a
late dark energy era, in the presence of a light axion particle
which plays the role of the dark matter component of the Universe.
Particularly, the early-time phenomenology is dominated by an
$R^2$ term, while the presence of the other terms $f(R)$ ensure
the occurrence of the early and late-time dark energy eras. The
inflationary phenomenology is compatible with the Planck 2018 data
for inflation, while the late-time dark energy era is compatible
with the Planck 2018 constraints on the cosmological parameters.
Also, the model exhibits an early dark energy era, at $z\sim 2.5$
approximately, followed by a deceleration era, which starts at
approximately $z\sim 1.5$, which in turn is followed by a
late-time dark energy era for redshifts $z\sim 0.5$, which lasts
for approximately 5 billion years  up to present time. A notable
feature of our model is that the dark energy era is free from dark
energy oscillations, at least in the redshift interval $z=[0,10]$.
In addition, we also discuss several features related to
observational data at $z\sim 2.34$, at which redshift intricate
observational data exist in the literature. Moreover, the
numerical code for the dark energy phenomenology, written in
Python 3, is presented in the end of the article. Finally, the
model has another interesting characteristic, a sudden jump of the
value of the Hubble rate in the redshift interval $z\sim [2,2.6]$
where its value suddenly increases and then decreases until $z\sim
0$.
\end{abstract}


\maketitle

\section{Introduction}

The observation that utterly changed the way of thinking in modern
theoretical cosmology, was made in the late 90's
\cite{Riess:1998cb} and indicated that the Universe is expanding
in an accelerated way, contrary to the standard model of
cosmology. This late-time acceleration era is known as the dark
energy era, and to date still remains a mystery on its driving
force. In the context of General Relativity (GR), the dark energy
era can be generated by using a cosmological constant, and the
model that is successful in fitting the observational data is the
$\Lambda$-Cold-Dark-Matter ($\Lambda$CDM) model. However, the
$\Lambda$CDM model strongly relies on the presence of a tiny
cosmological constant, which is added by hand in the model, and of
course in the presence of the other mysterious component of the
dark sector, dark matter
\cite{Bertone:2004pz,Bergstrom:2000pn,Mambrini:2015sia,Profumo:2013yn,Hooper:2007qk,Oikonomou:2006mh}.
The fact that the cosmological constant has such a small value, is
in conflict with the vacuum energy predictions of quantum field
theory, thus such a shortcoming has to be alleviated in a
theoretical way. Moreover, in the context of GR, the early and
late time era are usually described by two distinct theories, in
the case of inflation, the usual description is given by a
canonical scalar field with a sufficiently flat potential, while
the late-time era is generated by the cosmological constant. Thus
a unified description is lacking in the context of GR. Modified
gravity bridges the gap between the early and late-time
phenomenology, since a unified description can be achieved, see
for example the pioneer article \cite{Nojiri:2003ft} and Refs.
\cite{Nojiri:2007as,Nojiri:2007cq,Cognola:2007zu,Nojiri:2006gh,Appleby:2007vb,Elizalde:2010ts,Odintsov:2020nwm,Sa:2020fvn}
and also Refs.
\cite{reviews1,reviews2,reviews3,reviews4,reviews5,reviews6} for
reviews. Apart from the unified description of the early and the
late-times acceleration eras, modified gravity has the attribute
of generating a dark energy era with non-constant equation of
state (EoS) parameter, since for the $\Lambda$CDM model, the dark
energy era has the EoS $P=-\rho$.

In this paper, we shall present an $F(R)$ gravity model for which
the unification of the early and the late-time acceleration eras
can be achieved. In addition, the model also generates an early
dark energy era around $z\sim 5$, followed by a deceleration era,
and the latter is followed by the present day acceleration. In
addition to the $F(R)$ gravity model, we assume the presence of a
light misalignment axion field
\cite{Preskill:1982cy,Abbott:1982af,Dine:1982ah,Marsh:2015xka,Sikivie:2006ni,Raffelt:2006cw,Linde:1991km,Co:2019jts,Marsh:2017yvc,Odintsov:2019mlf,Nojiri:2019nar,Nojiri:2019riz,Odintsov:2019evb,Cicoli:2019ulk,Fukunaga:2019unq,Caputo:2019joi,maxim,Chang:2018rso,Irastorza:2018dyq,Anastassopoulos:2017ftl,Sikivie:2014lha,Sikivie:2010bq,Sikivie:2009qn,Caputo:2019tms,Masaki:2019ggg,Soda:2017sce,Soda:2017dsu,Aoki:2017ehb,Masaki:2017aea,Aoki:2016kwl,Obata:2016xcr,Aoki:2016mtn,Ikeda:2019fvj,Arvanitaki:2019rax,Arvanitaki:2016qwi,Arvanitaki:2014wva,Arvanitaki:2014dfa,Sen:2018cjt,Cardoso:2018tly,Rosa:2017ury,Yoshino:2013ofa,Machado:2019xuc,Korochkin:2019qpe,Chou:2019enw,Chang:2019tvx,Crisosto:2019fcj,Choi:2019jwx,Kavic:2019cgk,Blas:2019qqp,Guerra:2019srj,Tenkanen:2019xzn,Huang:2019rmc,Croon:2019iuh,Day:2019bbh,Odintsov:2020iui,Nojiri:2020pqr,Odintsov:2020nwm}
which has a broken Peccei-Quinn symmetry during inflation. During
inflation the axion remains frozen at its vacuum expectation
value, but after the inflationary era it starts oscillating and
behaves as cold dark matter. We analyze the inflationary
phenomenology of the model at hand in detail and we demonstrate
that it is compatible with the Planck constraints on inflation
\cite{Akrami:2018odb}. Also we provide a thorough numerical
analysis of the dark energy era, utilizing a numerical code
written in Python 3 which is freely available and can be found in
\cite{code}. For the dark energy era we quantify our analysis by
expressing all the physical quantities as functions of the
redshift, and as functions of an appropriate statefinder quantity.
As we demonstrate, the results are compatible with the latest
Planck constraints on the cosmological parameters
\cite{Aghanim:2018eyx}. Furthermore, an early dark energy era is
produced by our theoretical model, starting approximately at
$z\sim 2.5$ and ending at $z\sim 1.5$. Accordingly, the model
decelerates until $z\sim 0.5$ and after that it accelerates again
until the present time era. In the literature, several models of
early dark energy models have been worked out, see for example
Refs.
\cite{Doran:2006kp,Bhattacharyya:2019lvg,Sakstein:2019fmf,Tian:2019enx,Nojiri:2019fft}.
Furthermore, we analyze an interesting feature of the model, a
sudden jump in the Hubble rate at $z\sim 2$. Interestingly enough,
the Hubble rate at the aforementioned redshift, has a local
minimum followed by a local maximum, and after that it decreases
until the present day. Finally, we briefly discuss our results in
view of the observational data \cite{Delubac:2014aqe} for
redshifts $z\sim 2$.

The motivation for using a combined scalar field and $f(R)$
gravity framework is two-fold, firstly the scalar field in our
case is basically a misalignment axion field, which is a
primordial scalar field coming as a remnant of the quantum theory
preceding the inflationary era. Secondly, the inflationary era
emerged directly from a quantum epoch of our Universe, where all
interactions were unified and possibly spacetime could be higher
dimensional. However, during the inflationary era, the Universe
was four dimensional and classical, nevertheless there
inflationary effective Lagrangian may contain higher order
curvature terms and even string-corrections as remnants of the
quantum era of our Universe. Such terms could be $f(R)$ gravity
terms, such as the case we shall study in this paper, or even
Gauss-Bonnet terms coupled to the scalar field, thus effectively
having an Einstein-Gauss-Bonnet theory
\cite{Hwang:2005hb,Nojiri:2006je,Cognola:2006sp,Nojiri:2005vv,Nojiri:2005jg,Satoh:2007gn,Bamba:2014zoa,Yi:2018gse,Guo:2009uk,Guo:2010jr,Jiang:2013gza,Kanti:2015pda,vandeBruck:2017voa,Kanti:1998jd,Pozdeeva:2020apf,Fomin:2020hfh,DeLaurentis:2015fea,Chervon:2019sey,Nozari:2017rta,Odintsov:2018zhw,Kawai:1998ab,Yi:2018dhl,vandeBruck:2016xvt,Kleihaus:2019rbg,Bakopoulos:2019tvc,Maeda:2011zn,Bakopoulos:2020dfg,Ai:2020peo,Odintsov:2019clh,Oikonomou:2020oil,Odintsov:2020xji,Oikonomou:2020sij,Odintsov:2020zkl,Odintsov:2020sqy,Odintsov:2020mkz,Easther:1996yd,Antoniadis:1993jc,Antoniadis:1990uu,Kanti:1995vq,Kanti:1997br}.
Einstein-Gauss-Bonnet theories in the presence of $R^2$ terms were
studied in Ref. \cite{Odintsov:2020ilr}, and it would be
interesting to study these in the case that the scalar field is
actually the axion, however this is out of the scopes of this
paper. Also let us notice that if higher order curvature terms are
present in the inflationary effective Lagrangian, these terms
would apparently dominate the evolution at early times compared to
the Einstein-Hilbert term or even the frozen axion, since even for
the low scale inflationary scenario, the Hubble rate during
inflation is of the order of $H_I\sim \mathcal{O}(10^{13})$GeV. We
shall explicitly show this in the following sections.

\section{Essential Features of the $f(R)$-Axion Model}

We shall assume that the gravitational action of our model has the
following form,
\begin{equation}
\label{mainaction} \mathcal{S}=\int d^4x\sqrt{-g}\left[
\frac{1}{2\kappa^2}F(R)-\frac{1}{2}\partial^{\mu}\phi\partial_{\mu}\phi-V(\phi)+\mathcal{L}_m
\right]\, ,
\end{equation}
with $\kappa^2=\frac{1}{8\pi G}=\frac{1}{M_p^2}$, where $G$ is
Newton's gravitational constant and $M_p$ is the reduced Planck
mass. In addition, $\mathcal{L}_m$ describes the Lagrangian
density of the perfect matter fluids present, which in our case is
radiation. The light axion field is quantified by the scalar field
$\phi$ and its dynamics will be explained shortly. The $F(R)$
gravity model has the following form,
\begin{equation}\label{starobinsky}
F(R)=R+\frac{1}{M^2}R^2+\lambda  R \exp \left(-\frac{\gamma \Lambda
}{R}\right)+\gamma \lambda  \Lambda -\frac{\Lambda
\left(\frac{R}{m_s^2}\right)^{\delta}}{\zeta}\, ,
\end{equation}
where $m_s$ is $m_s^2=\frac{\kappa^2\rho_m^{(0)}}{3}$,
$\rho_m^{(0)}$ is the present day energy density of cold dark
matter, and $0<\delta<1$, while $\zeta$ and $\gamma$ are dimensionless constants to be specified latter. Also, $M$ is chosen to be $M= 1.5\times
10^{-5}\left(\frac{N}{50}\right)^{-1}M_p$ for inflationary
phenomenological reasoning \cite{Appleby:2009uf}, where $N$ is the
$e$-foldings number. Moreover, the parameter $\Lambda$ takes
values of the order of the cosmological constant and has
dimensions eV$^2$, while $\lambda$ is a dimensionless parameter.
For a flat Friedmann-Robertson-Walker (FRW) metric,
\begin{equation}
\label{metricfrw} ds^2 = - dt^2 + a(t)^2 \sum_{i=1,2,3}
\left(dx^i\right)^2\, ,
\end{equation}
the field equations for the $f(R)$ gravity in the presence of
axion dark matter and radiation, and the scalar field read,
\begin{align}\label{eqnsofmkotion}
& 3 H^2F_R=\frac{RF_R-F}{2}-3H\dot{F}_R+\kappa^2\left(
\rho_r+\frac{1}{2}\dot{\phi}^2+V(\phi)\right)\, ,\\ \notag &
-2\dot{H}F=\kappa^2\dot{\phi}^2+\ddot{F}_R-H\dot{F}_R
+\frac{4\kappa^2}{3}\rho_r\, ,
\end{align}
\begin{equation}\label{scalareqnofmotion}
\ddot{\phi}+3H\dot{\phi}+V'(\phi)=0
\end{equation}
with $F_R=\frac{\partial F}{\partial R}$, and the``dot'' and the
``prime'' denote differentiation with respect to the cosmic time
and the scalar field respectively.

Let us turn our focus to the axion scalar field, and let us
discuss how it evolves during the evolution of the Universe. Here
we shall briefly outline the axion dynamics, which is described in
detail in Ref. \cite{Odintsov:2020nwm,Marsh:2015xka}. We shall
consider an axion scalar field with a broken $U(1)$ Peccei-Quinn
symmetry during the inflationary era, with a potential during
inflation of the form \cite{Marsh:2015xka},
\begin{equation}\label{axionpotential}
V(\phi )\simeq \frac{1}{2}m_a^2\phi^2_i\, ,
\end{equation}
where $\phi_i$ is the large vacuum expectation value of the axion
field during inflation. The axion field is frozen during the
inflationary era to its vacuum expectation value and does not
evolve dynamically, so the following initial conditions describe
it \cite{Marsh:2015xka},
\begin{equation}\label{axioninitialconditions}
\ddot{\phi}(t_i)\simeq 0,\,\,\,\dot{\phi}(t_i)\simeq
0,\,\,\,\phi(t_i)\equiv\phi_i=f_a\theta_a\, ,
\end{equation}
where $t_i$ is the cosmic time during the inflationary era, $f_a$
is the axion decay constant, $\theta_a$ denotes the initial
misalignment angle, and $m_a$ is the axion mass. The above frozen
dynamical evolution continues to occur as long as $H\gg m_a$. When
$m_a\sim H$, the axion starts to oscillate and thus for the eras
that $m_a\succeq H$, the axion energy density evolves as
\cite{Odintsov:2020nwm,Marsh:2015xka}
\begin{equation}\label{axionenergydensityfinalpost}
\rho_a\simeq \rho_m^{(0)}a^{-3}\, ,
\end{equation}
where $\rho_m^{(0)}=\frac{1}{2}m_a^2\phi_i^2$, hence it evolves as
a cold dark matter component for the eras for which the condition
$m_a\succeq H$ holds true. In the following we shall study the
dynamics of the model (\ref{mainaction}) for both the early and
late time eras.

\section{Inflationary Evolution of the Model}

During the inflationary era, the Hubble rate is of the order of
$H_I=10^{13}$GeV for most Grand Unified Theories with low-scale
inflation. Thus the Ricci scalar $R\sim H_I^2$ takes quite large
value, hence at leading order, the effective $F(R)$ of the model
(\ref{starobinsky}) becomes,
\begin{equation}\label{effectivelagrangian1}
F(R)\simeq (1+\lambda) R+\frac{R^2}{M^2}-\frac{\gamma^3 \lambda
\Lambda^3}{6 R^2}+\frac{\gamma^2 \lambda  \Lambda^2}{2
R}-\frac{\Lambda  \left(\frac{R}{m_s^2}\right)^{\delta }}{\zeta
}\, ,
\end{equation}
and thus at leading order, the effective $F(R)$ gravity during
inflation is equal to,
\begin{equation}\label{effectivelagrangian1}
F(R)\simeq (1+\lambda)R+\frac{1}{M^2}R^2-\frac{\Lambda
\left(\frac{R}{m_s^2}\right)^{\delta}}{\zeta}\, .
\end{equation}
Now in order to analyze the inflationary dynamics, we need to
understand which terms in the equations of motion
(\ref{eqnsofmkotion}) dominate the evolution. Firstly, $m_s$ which
was defined below Eq. (\ref{starobinsky}) is $m_s^2\simeq
1.87101\times 10^{-67}$eV$^2$ and the parameter $M$ coupled to the
$R^2$ term in Eq. (\ref{starobinsky}) is $M= 1.5\times
10^{-5}\left(\frac{N}{50}\right)^{-1}M_p$ \cite{Appleby:2009uf},
hence for $N\sim 60$, $M$ is equal to $M\simeq 3.04375\times
10^{22}$eV. Moreover, since the inflationary era is a slow-roll
era, meaning that $\dot{H}\ll H^2$, we have $R\simeq 12 H^2$, and
for $H= H_I\sim 10^{13}$GeV, we have $R\sim 1.2\times
10^{45}$eV$^2$. Also the reduced Planck mass is $M_p\simeq 2.435
\times 10^{27}$eV, and the parameter $\Lambda$ will be assumed to
take the value $\Lambda\simeq 11.895\times 10^{-67}$eV$^2$, which
is close to the value of the cosmological constant at present day.
Finally, for phenomenological reasoning, the values of $\phi_i$
and $m_a$ are chosen to be  $\phi_i=\mathcal{O}(10^{15})$GeV and
$m_a\simeq \mathcal{O}(10^{-14})$eV. Let us now compare the terms
appearing in the equations of motion, having in mind that the
leading order terms in the $F(R)$ gravity are $\sim R$ and $R^2$.
Firstly, the radiation density term during inflation is highly
subdominant since $\kappa^2\rho_r \sim e^{-N}$ during inflation,
and also the kinetic term of the axion scalar $\sim
\dot{\phi}_i^2$ is also subdominant, since the axion during
inflation is frozen and obeys the initial conditions
(\ref{axioninitialconditions}). Now let us proceed with the
potential term which is of the order $\kappa^2V(\phi_i)\sim
\mathcal{O}(8.41897\times 10^{-36})$eV$^{2}$, while the terms $R$
and $R^2$ are, $R\sim 1.2\times \mathcal{O}(10^{45})$eV$^2$ and
also $R^2/M^2\sim \mathcal{O}(1.55\times 10^{45})$eV$^2$. Finally
the power-law curvature terms, for $\delta=0.1$ and $\zeta=0.2$
(which are the values for $\zeta$ and $\delta$ we shall also
assume for the late-time analysis) take values of the order
$\frac{\Lambda \left(\frac{R}{m_s}\right)^{0.1}}{0.2}\sim
 \mathcal{O}(10^{-55})$eV$^2$ and the rest of the terms are highly
subdominant. Thus during the inflationary era, the dynamical
evolution of the cosmological system is controlled by the $f(R)$
gravity,
\begin{equation}\label{effectivelagrangian2}
F(R)\simeq (1+\lambda)R+\frac{1}{M^2}R^2\, .
\end{equation}
By looking the effective form of the $F(R)$ gravity appearing in
Eq. (\ref{effectivelagrangian2}), one would expect that the term
$(1+\lambda)$ will affect the phenomenology of this deformed $R^2$
model. We shall work out in detail the inflationary phenomenology
of the model (\ref{effectivelagrangian2}) and to our surprise, for
the $R^2$ model there is no difference between the $\lambda \neq
0$ and $\lambda=0$ theories. Let us explicitly show this, so by
substituting Eq. (\ref{effectivelagrangian2}) in the Friedmann
equation in Eq. (\ref{eqnsofmkotion}), we get without any
approximation, the following differential equation,
\begin{equation}\label{diffeqndefomedstaro}
\ddot{H}+3 H \dot{H}-\frac{\dot{H}^2}{2 H}+\frac{1}{12} \lambda
M^2 H+\frac{1}{12} M^2 H=0\, .
\end{equation}
Since the inflationary era is governed by the slow-roll
approximation, quantified by the relations,
\begin{equation}\label{slowrollapproximation}
\ddot{H}\ll H \dot{H},\,\,\,\dot{H}\ll H^2\, ,
\end{equation}
in view of the relations (\ref{slowrollapproximation}), the
Friedmann equation (\ref{diffeqndefomedstaro}) becomes,
\begin{equation}\label{finalformfriedmanneqn}
\dot{H}\simeq -\frac{1}{36} (\lambda +1) M^2\, ,
\end{equation}
which is solve easily and yields a quasi-de Sitter evolution,
\begin{equation}\label{quasidesitter}
H(t)=H_I-\frac{1}{36} t \left(\lambda  M^2+M^2\right)\, ,
\end{equation}
where $H_I$ is an integration constant, which is basically the
inflationary scale. Now we can obtain easily the phenomenology of
the model and the slow-roll parameters are
\cite{Hwang:2005hb,reviews1,Odintsov:2020thl},
\begin{equation}
\label{restofparametersfr}\epsilon_1=-\frac{\dot{H}}{H^2}, \quad
\epsilon_2=0\, ,\quad \epsilon_3= \frac{\dot{F}_R}{2HF_R}\, ,\quad
\epsilon_4=\frac{\ddot{F}_R}{H\dot{F}_R}\,
 ,
\end{equation}
and the spectral index of the primordial scalar curvature
perturbations and the tensor-to-scalar ratio are
\cite{Odintsov:2020thl},
\begin{equation}
\label{spectralfinal} n_s\simeq 1-6\epsilon_1-2\epsilon_4\, ,
\end{equation}
\begin{equation}
\label{tensorfinal} r\simeq 48\epsilon_1^2\, .
\end{equation}
Now, the exact expression for the slow-roll index can easily be
found \cite{Odintsov:2020thl}, and it is equal to,
\begin{equation}\label{finalapproxepsilon4}
\epsilon_4\simeq -\frac{24
F_{RRR}H^2}{F_{RR}}\epsilon_1-\epsilon_1\, ,
\end{equation}
hence for the deformed $R^2$ model of Eq.
(\ref{effectivelagrangian2}), the slow-roll index $\epsilon_4$ is
$\epsilon_4\simeq -\epsilon_1$, hence the spectral index of the
primordial curvature perturbations and the tensor-to-scalar ratio
are,
\begin{equation}
\label{spectralfinal} n_s\simeq 1-4\epsilon_1\, ,
\end{equation}
\begin{equation}
\label{tensorfinal} r\simeq 48\epsilon_1^2\, .
\end{equation}
The slow-roll index $\epsilon_1$ can easily be found by using the
analytic quasi-de Sitter solution for the Hubble rate
(\ref{finalformfriedmanneqn}), and it is equal to,
\begin{equation}\label{epsilon1indexanalytic}
\epsilon_1=-\frac{-\lambda  M^2-M^2}{36 \left(H_I-\frac{1}{36} t
\left(\lambda M^2+M^2\right)\right)^2}\, ,
\end{equation}
and from it we can easily find the time instance at which
inflation ends, namely, $t_f$, by solving the equation
$\epsilon_1(t_f)=1$, and the solution is,
\begin{equation}\label{finaltimeinstance}
t_f=\frac{6 \left(6 H_I \lambda  M^2+6 H_I M^2-\sqrt{\lambda ^3
M^6+3 \lambda ^2 M^6+3 \lambda M^6+M^6}\right)}{\lambda ^2 M^4+2
\lambda  M^4+M^4}\, .
\end{equation}
We can now use the definition of the $e$-foldings number $N$,
\begin{equation}\label{efoldingsnumber}
N=\int_{t_i}^{t_f}H(t)dt\, ,
\end{equation}
in order to find $t_i$ as a function of the $e$-foldings number
and the rest of the parameters, so we have,
\begin{equation}\label{ti}
t_i=\frac{6 \left(6 H_I+\sqrt{(\lambda +1) M^2 (2
N+1)}\right)}{(\lambda +1) M^2}\, .
\end{equation}
Now if we calculate the slow-roll index $\epsilon_1$ at the first
horizon crossing time instance $t_i$, we get remarkably a
$\lambda$-independent result,
\begin{equation}\label{epsilon1lambdaind}
\epsilon_1(t_i)=\frac{1}{1+2N}\, ,
\end{equation}
so the spectral index and the tensor-to-scalar ratio are at
leading order $n_s\sim 1-\frac{2}{N}$ and $r\sim \frac{12}{N^2}$,
which are identical to the $R^2$ model with $\lambda=0$. The model
is compatible with the latest Planck data \cite{Aghanim:2018eyx}.
Thus the inflationary phenomenology of the model
(\ref{starobinsky}) is viable and compatible with the Planck data.
As a final comment, the fact that the parameter $\lambda$ does not
affect at all the phenomenology of the model, is an artifact of
the $R^2$ model and of the fact that the axion scalar is not
dynamically evolving during the inflationary era. In fact it can
be shown that if a dynamically evolving canonical scalar field is
present, this rescaled Einstein-Hilbert $R$ term affects the
inflationary phenomenology significantly in some cases
\cite{workinprogress}.

\section{Early and Late Dark Energy Eras}

For the $F(R)$ gravity theory in the presence of radiation and the
axion dark matter fluid, the field equations for the flat FRW
metric can be cast in the Einstein-Hilbert form as follows,
\begin{align}\label{flat}
& 3H^2=\kappa^2\rho_{tot}\, ,\\ \notag &
-2\dot{H}=\kappa^2(\rho_{tot}+P_{tot})\, ,
\end{align}
where $\rho_{tot}=\rho_{a}+\rho_{DE}+\rho_r$ stands for the total
energy density, $\rho_a$ is the axion field energy density, which
recall that it scales as $\sim a^{-3}$, while $\rho_r$ is the
radiation energy density and $\rho_{DE}$ is the dark energy
density, a purely geometric term since it is equal to,
\begin{equation}\label{degeometricfluid}
\kappa^2\rho_{DE}=\frac{F_R R-F}{2}+3H^2(1-F_R)-3H\dot{F}_R\, .
\end{equation}
Also $P_{tot}=P_r+P_{a}+P_{DE}$ denotes the total pressure of the
cosmological fluid, and the dark energy pressure is,
\begin{equation}\label{pressuregeometry}
\kappa^2P_{DE}=\ddot{F}_R-H\dot{F}_R+2\dot{H}(F_R-1)-\kappa^2\rho_{DE}\,
.
\end{equation}
In this section we aim to study the late-time behavior of the
model (\ref{starobinsky}), by solving numerically the Friedmann
equation (\ref{eqnsofmkotion}). To this end, we shall use the
redshift as a dynamical parameter quantifying the evolution
instead of the cosmic time, and also we shall introduce an
appropriate statefinder quantity, which will make the dark energy
effects more transparent. The redshift $z$ is defined as,
\begin{equation}\label{redshift}
1+z=\frac{1}{a}\, ,
\end{equation}
and the present time scale factor, that is at $z=0$, is assumed to
be equal to unity. The statefinder we shall use is the function
$y_H(z)$
\cite{Hu:2007nk,Bamba:2012qi,Odintsov:2020vjb,Odintsov:2020nwm,Odintsov:2020qyw,reviews1},
\begin{equation}\label{yHdefinition}
y_H(z)=\frac{\rho_{DE}}{\rho^{(0)}_m}\, ,
\end{equation}
with $\rho^{(0)}_m$ being the present time energy density of cold
dark matter. Obviously, $y_H(z)$ is a dark energy dependent
quantity, and it is different from zero, when geometric terms
appear in the gravitational action. We can write the first
Friedmann equation in terms of the statefinder quantity $y_H(z)$
and by recalling that for the axion field $\rho_a=\rho^{(0)}_m
(1+z)^3$,
\begin{equation}\label{finalexpressionyHz}
y_H(z)=\frac{H^2}{m_s^2}-(1+z)^{3}-\chi (1+z)^4\, ,
\end{equation}
where we introduced the parameter
$\chi=\frac{\rho^{(0)}_r}{\rho^{(0)}_m}\simeq 3.1\times 10^{-4}$,
and $\rho^{(0)}_r$ is the radiation energy density, and recall
that
$m_s^2=\frac{\kappa^2\rho^{(0)}_m}{3}=H_0\Omega_c=1.37201\times
10^{-67}$eV$^2$. Apparently the statefinder $y_H(z)$ clearly shows
deviations from the standard model of cosmology, and it is
constant for the $\Lambda$CDM model. Hence it is indeed a
statefinder for dark energy since it shows deviations from the
Einstein gravity and also shows if the dark energy is dynamical or
not, and the latter issue is still a mystery too, along with dark
energy itself. In terms of the statefinder $y_H(z)$, the Friedmann
equation reads \cite{Bamba:2012qi},
\begin{equation}\label{differentialequationmain}
\frac{d^2y_H(z)}{d z^2}+J_1\frac{d y_H(z)}{d z}+J_2y_H(z)+J_3=0\,
,
\end{equation}
with the dimensionless functions $J_1$, $J_2$ and $J_3$ being
defined in the following way,
\begin{align}\label{diffequation}
& J_1=\frac{1}{z+1}\left(
-3-\frac{1-F_R}{\left(y_H(z)+(z+1)^3+\chi (1+z)^4\right) 6
m_s^2F_{RR}} \right)\, , \\ \notag & J_2=\frac{1}{(z+1)^2}\left(
\frac{2-F_R}{\left(y_H(z)+(z+1)^3+\chi (1+z)^4\right) 3
m_s^2F_{RR}} \right)\, ,\\ \notag & J_3=-3(z+1)-\frac{\left(1-F_R
\right)\Big{(}(z+1)^3+2\chi (1+z)^4
\Big{)}+\frac{R-F}{3m_s^2}}{(1+z)^2\Big{(}y_H(z)+(1+z)^3+\chi(1+z)^4\Big{)}6m_s^2F_{RR}}\,
,
\end{align}
with $F_{RR}=\frac{\partial^2 F}{\partial R^2}$. Our aim is to
solve numerically the differential equation
(\ref{differentialequationmain}) in the redshift interval
$z=[0,10]$, by using appropriate physical motivated initial
conditions which correspond to the late stages of the matter
domination era. These are,
\begin{equation}\label{generalinitialconditions}
y_H(z_f)=\frac{\Lambda}{3m_s^2}\left(
1+\frac{(1+z_f)}{1000}\right)\, , \,\,\,\frac{d y_H(z)}{d
z}\Big{|}_{z=z_f}=\frac{1}{1000}\frac{\Lambda}{3m_s^2}\, .
\end{equation}
We developed a numerical code appropriately constructed to
integrate the differential equation
(\ref{differentialequationmain}) backwards from $z=10$ to $z=0$,
using PYTHON 3, and specifically the $``\mathrm{solve}_{\_
}\mathrm{ivp}''$ function of the SCIPY module. Also we performed
the analysis with Mathematica 11, and the results almost coincide
with the PYTHON outcomes. For the PYTHON code we used several
methods of numerical integration, like LSODA, BDF, RK45 (4th order
Runge-Kutta with variable step and dynamically 5th order), and
also the Radau, and the numerical results were similar in all
cases. The numerical code along with a pedagogical description of
the code and the physics of dark energy phenomenology, can be
found here \cite{code}.

Before presenting the results of the numerical analysis, let us
quote the functional forms of several quantities of interest, as
functions of the statefinder $y_H(z)$. The curvature as a function
of $y_H(z)$ reads,
\begin{equation}\label{ricciscalarasfunctionofz}
R(z)=3m_s^2\left( 4y_H(z)-(z+1)\frac{d y_H(z)}{d
z}+(z+1)^3\right)\, .
\end{equation}
The dark energy density parameter $\Omega_{DE}$ reads,
\begin{equation}\label{omegaglarge}
\Omega_{DE}(z)=\frac{y_H(z)}{y_H(z)+(z+1)^3+\chi (z+1)^4}\, .
\end{equation}
while the dark energy EoS parameter reads,
\begin{equation}\label{omegade}
\omega_{DE}(z)=-1+\frac{1}{3}(z+1)\frac{1}{y_H(z)}\frac{d
y_H(z)}{d z}\, ,
\end{equation}
and the total EoS parameter is,
\begin{equation}\label{totaleosparameter}
\omega_{tot}(z)=\frac{2 (z+1) H'(z)}{3 H(z)}-1\, .
\end{equation}
Finally, the deceleration parameter reads,
\begin{align}\label{statefinders}
& q=-1-\frac{\dot{H}}{H^2}=-1+(z+1)\frac{H'(z)}{H(z)}\, .
\end{align}
Also for $\Lambda$CDM model, the Hubble rate equals to,
\begin{equation}\label{lambdacdmhubblerate}
H_{\Lambda}(z)=H_0\sqrt{\Omega_{\Lambda}+\Omega_M(z+1)^3+\Omega_r(1+z)^4}\,
,
\end{equation}
with $H_0$ being the value of the Hubble rate at present time,
which according to the latest Planck data is $H_0\simeq
1.37187\times 10^{-33}$eV according to the latest Planck data
\cite{Aghanim:2018eyx}. Also $\Omega_{\Lambda}\simeq 0.681369$ and
$\Omega_M\sim 0.3153$ \cite{Aghanim:2018eyx}, while
$\Omega_r/\Omega_M\simeq \chi$, with $\chi$ being presented below
Eq. (\ref{finalexpressionyHz}). Also for the numerical analysis we
shall choose $\delta=0.1$, $\lambda=0.999$, $\zeta=0.2$ and
$\gamma=5.1$ in Eq. (\ref{starobinsky}).

\begin{figure}
\centering
\includegraphics[width=18pc]{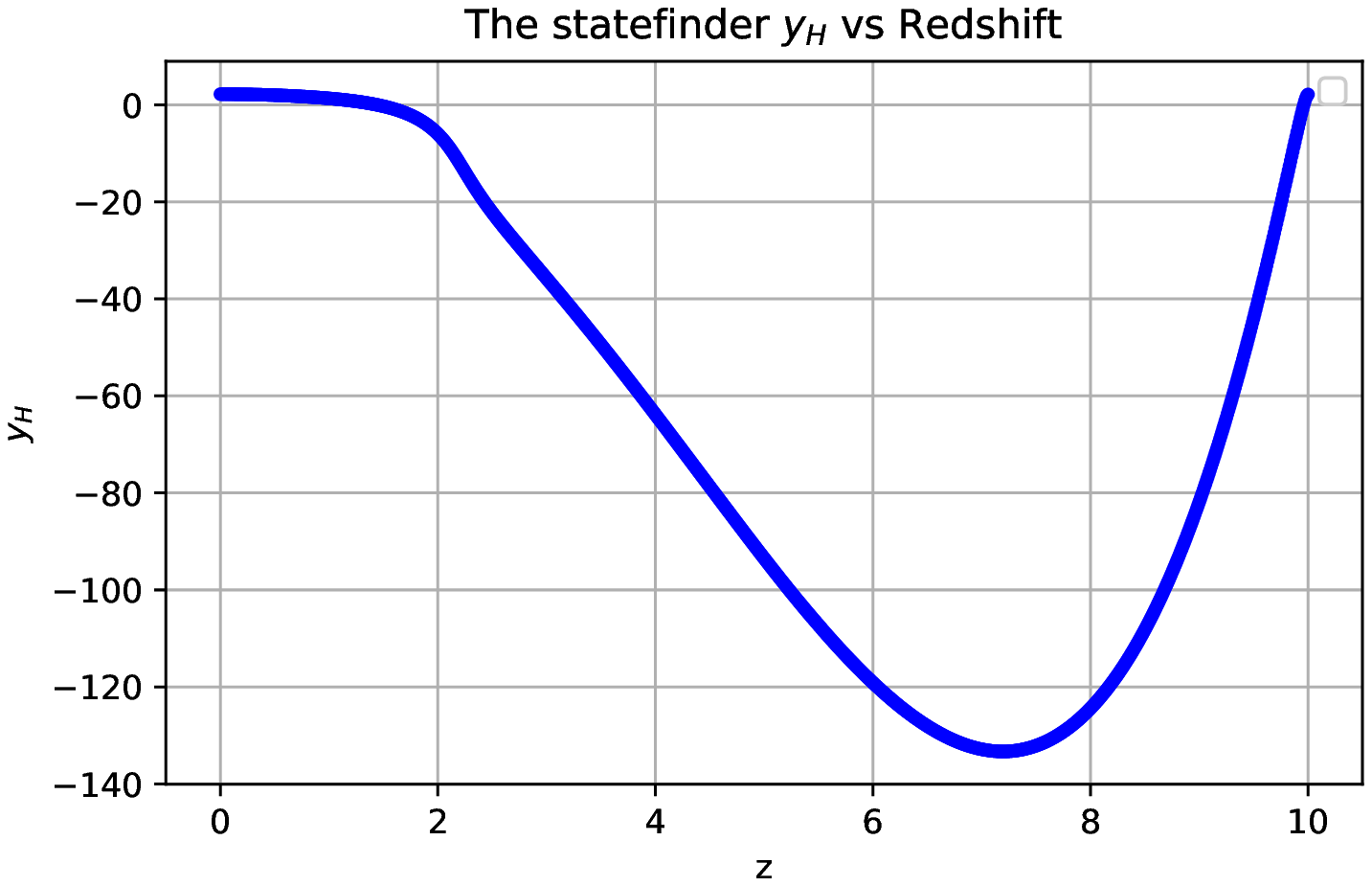}
\includegraphics[width=18pc]{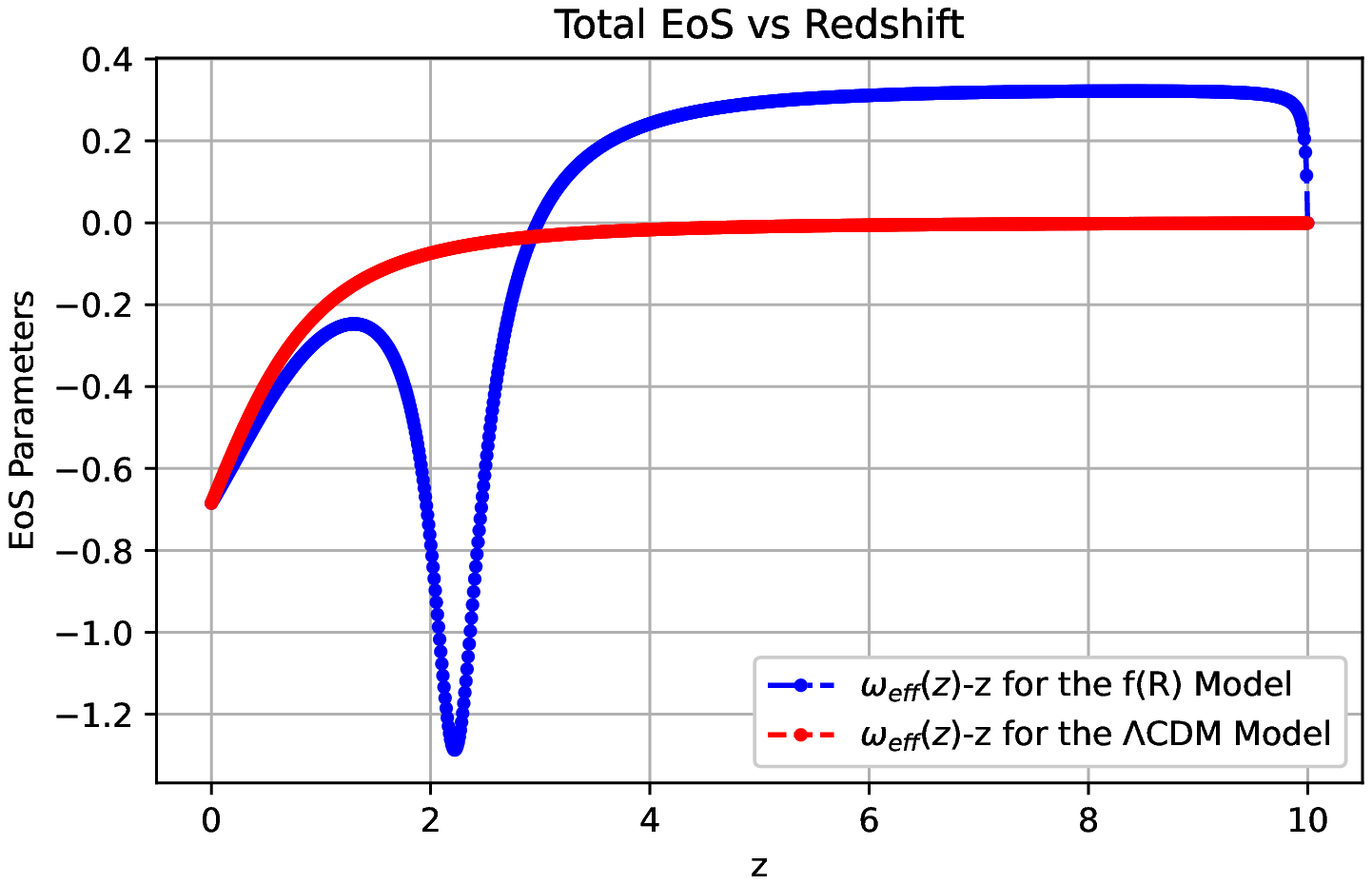}
\caption{The statefinder function $y_H$ for geometric dark energy
as a function of the redshift (left plot), and the total EoS
parameter $\omega_{tot}$ (right plot).}\label{plot1}
\end{figure}

\begin{figure}
\centering
\includegraphics[width=18pc]{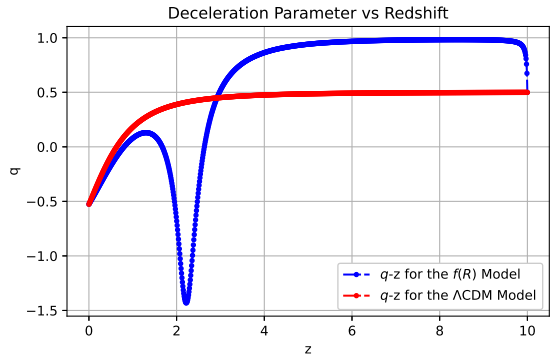}
\caption{The deceleration parameter, as a function of the
redshift.} \label{plot1a}
\end{figure}

\begin{figure}[h!]
\centering
\includegraphics[width=18pc]{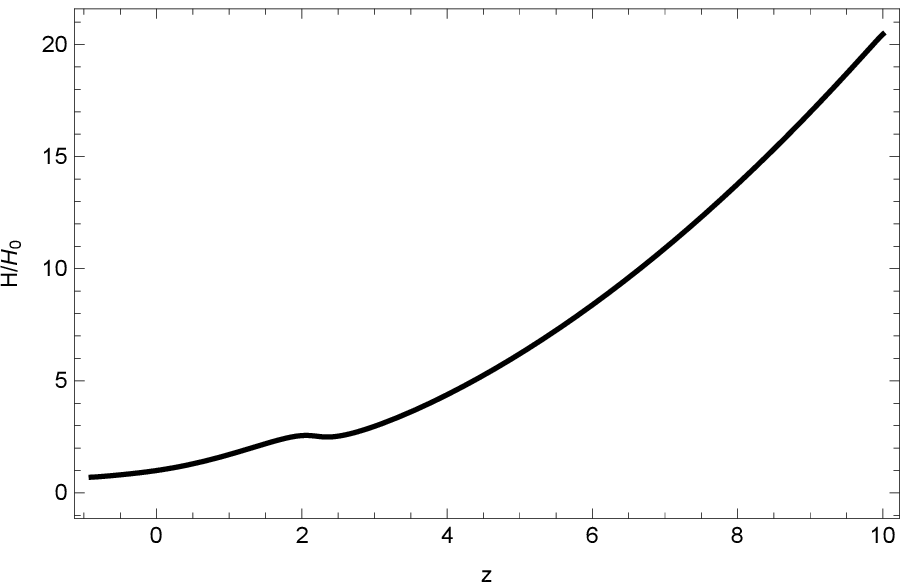}
\includegraphics[width=18pc]{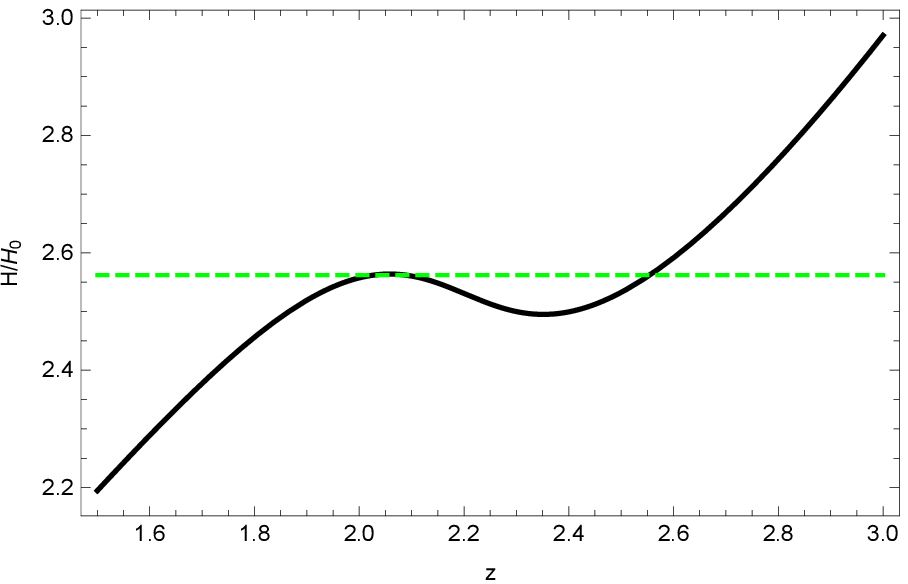}
\caption{$\frac{H(z)}{H(0)}$ as a function of the redshift. In the
right plot the intriguing behavior of the Hubble rate for
redshifts $z\sim 2.6 $ until $z\sim 2$ is shown in a more
transparent way. The green line in the right plot corresponds to
the $H/H(0)$ value $2.56196$.} \label{plot2}
\end{figure}

Now let us discuss the results of our numerical analysis and we
gather the most characteristic examples of the late-time behavior
for the model (\ref{starobinsky}) in Fig. \ref{plot1}, where we
plot the statefinder $y_H(z)$ (left plot), the total EoS parameter
(right plot) and in Fig. \ref{plot1a} the deceleration parameter,
as functions of the redshift. Also in Fig. \ref{plot2} we plot the
Hubble rate as a function of the redshift. The behavior of the
statefinder $y_{H}(z)$ clearly shows that $y_H(z)$ is essentially
negative for the whole interval $z=[0,10]$, and this has
interesting consequences in view of the observational data
reported in Ref. \cite{Delubac:2014aqe}, so we comment on the
$y_H(z)$ behavior in the next section. The behavior of the
deceleration parameter can clearly indicate when the Universe is
accelerating and when the Universe is decelerating. In our case,
the Universe decelerates until $z\sim 2.5$, then it enters an
acceleration phase, until $z\sim 1.5$, followed by a decelerating
epoch which lasts for nearly 4 billion years until $z\sim 0.5$.
After $z\sim 0.5$ the Universe enters the final acceleration era
which lasts until present day, nearly 5 billion years. In Fig.
\ref{plot1a}, we also quote the $\Lambda$CDM behavior of the
deceleration parameter (red), and as it can be seen, the $F(R)$
model is indistinguishable from the $\Lambda$CDM model only for
the last two billion years approximately, from $z\sim 0.2$ to
$z=0$. The same conclusion applies in the behavior of the total
EoS parameter in the right plot of Fig. \ref{plot1}, where the red
curve represents the behavior of the $\Lambda$CDM model. Thus the
$F(R)$ gravity model clearly has two dark energy epochs, one early
and one late dark energy epoch, and in between these to dark
energy eras, the Universe is decelerating. The same behavior can
be verified by looking the right plot of Fig. \ref{plot1}, that is
the total EoS parameter for the $F(R)$ gravity model (blue curve),
while the red curve corresponds to the $\Lambda$CDM model.
An interesting behavior in the Hubble rate which occurs for $z\sim
2.6 $ until $z\sim 2$, is that the Hubble rate at $z\sim 2.2$ has
a local minimum, followed by a local maximum at $z\sim 2$ and then
the Hubble rate evolves normally to its present day value for the
$F(R)$ gravity model. This intriguing behavior is presented in
Fig. \ref{plot2}, where we plot $H/H(0)$ as a function of the
redshift. The curious behavior is highlighted on the right plot,
and it seems that it is an inherent characteristic of the model,
we could not attribute this behavior to any other qualitative
reasoning. Also we should note that the fraction of the present
day value of the Hubble rate for the $F(R)$ gravity model, over
the Planck value for the Hubble rate at present day is
$\frac{H(0)}{H_0}=0.563636$.
Finally, let us discuss the viability of the $F(R)$ gravity model
in a more quantitative way, so let us give the values of the dark
energy density parameter $\Omega_{DE}$ and of the dark energy EoS
parameter at present day, and we compare these with the Planck
2018 constraints  on these quantities. The PYTHON code using the
``LSODA'' integration technique yields $\omega_{DE}(0)\simeq
-0.9961$ which is within the viability limits of the Planck
constraint $\omega_{DE}=-1.018\pm 0.031$, and accordingly the dark
energy density parameter is found equal to $\Omega_{DE}(0)\simeq
0.6872$, which is an acceptable value when compared to the Planck
constraint $\Omega_{DE}=0.6847\pm 0.0073$. Also the deceleration
parameter value at present day for the $F(R)$ gravity model is
$q(0)\simeq -0.5267$ while for the $\Lambda$CDM model is
$q=-0.527$, hence the two values are nearly identical.

A feature of the present $F(R)$ gravity model that is worth
mentioning is the fact that the model is free from dark energy
oscillations which are known to plague $F(R)$ gravity models for
the redshift interval $z=[5,10]$. This in an inherent
characteristic of the model as it proves, and thus we may conclude
that the occurrence of dark energy oscillations in $F(R)$ gravity
models is a model dependent feature.

In conclusion, the $F(R)$ gravity model (\ref{starobinsky}) in the
presence of a light axion field, produces a unified
phenomenological picture for the inflationary era and the dark
energy eras, with the intriguing characteristic of producing
actually two distinct dark energy era, one early and one late dark
energy eras, with a  brief deceleration epoch occurring between
these two deceleration eras. Finally let us finally comment that
the inverse integration PYTHON code we used and the Mathematica 11
results agree to a great extent.

\subsection{The Behavior at  $z\sim 2.34$}

Before closing, let us briefly discuss another interesting feature
of the model, which is related to the measurement of the Hubble
rate at $z=2.34$ reported by \cite{Delubac:2014aqe}. Assuming that
the observation is correct, we shall discuss how such an
observation is supported by the $F(R)$ gravity model
(\ref{starobinsky}). Basically, the observation of
\cite{Delubac:2014aqe}, indicates that the Hubble rate at redshift
$z\sim 2.34$ is $H(z=2.34)=222km/Mpc/sec$, a result which is also
discussed in
\cite{Moresco:2016mzx,Guo:2015gpa,Stern:2009ep,Chuang:2012qt}. An
interesting way to explain the result of \cite{Delubac:2014aqe},
is to make the assumption that $\rho_{DE}<0$ in Eq.
(\ref{finalexpressionyHz}), since otherwise one would get
$\Omega_{m}h^2=0.142$ which obviously does not agree with the
Cosmic Microwave Background related value of the Planck
Collaboration data $\Omega_c h^2=0.12\pm 0.001$
\cite{Aghanim:2018eyx}. Although that a negative dark energy
density sounds strange, this concept has appeared in the
literature \cite{Ahmed:2002mj,Cardenas:2014jya}. In our case,
negative dark energy density would mean negative values for the
statefinder $y_H(z)$, and obviously by looking the upper left plot
of Fig. \ref{plot1}, we can see that $y_H(z)$ is negative for
nearly the whole interval $z=[0,10]$. In fact it only becomes
positive at $z\sim 1.4$, where it has the value
$y_H(1.4)=0.227363$ and has the present day value
$y_H(0)=2.19527$. Let us also comment that negative values of
$y_H(z)$ in $F(R)$ gravity models frequently occur, since most
$f(R)$ gravity models are plagued with dark energy oscillations
for the interval $z=[5,10]$ \cite{Odintsov:2020nwm,reviews1}. The
present model is free from oscillations though, however $y_H$ is
negative for nearly the whole interval $z=[0,10]$.

\subsection{The Ghost Issue}

Before closing, we need to discuss the important issue of ghosts
in the essentially $f(R,\phi)$ theory we discussed in this paper.
The ghost instabilities may be developed in a modified gravity
theory if the wave speed of the cosmological perturbations, which
we will denote as $c_A$, is larger than unity. As was shown in
Ref. \cite{Hwang:2005hb} however, for $f(R,\phi)$ theories, the
wave speed of the scalar perturbations is unity, thus no ghost
instabilities occur in the curvature perturbations. In fact for a
flat spacetime, which is our case, the sound speed is equal to the
wave speed of the cosmological perturbations (see the last table
of Ref. \cite{Hwang:2005hb}). Let us elaborate on this issue a bit
further in order to clarify this important issue. We shall obtain
the cosmological perturbations by perturbing the flat FRW metric
$g_{\alpha \beta}^{(3)}$ in the following way \cite{Hwang:2005hb},
\begin{equation}\label{perturbationscosmo}
ds^2=-a^2(1+\alpha)d\eta^2-2\alpha^2\beta_{,\alpha}d \eta
dx^{\alpha}+a^2\left(g_{\alpha \beta}^{(3)}+2\varphi g_{\alpha
\beta}^{(3)}+2\gamma_{,\alpha | \beta}+2C_{\alpha \beta}
\right)dx^{\alpha}dx^{\beta}\, ,
\end{equation}
with $g_{\alpha \beta}^{(3)}$ being,
\begin{equation}\label{flatfrwanalytic}
g_{\alpha
\beta}^{(3)}dx^{\alpha}dx^{\beta}=dr^2+r^2(d\theta^2+\sin^2 \theta
d\phi^2)\, .
\end{equation}
In addition $\eta$ and $a$ in Eq. (\ref{perturbationscosmo})
denote the conformal time and the scale factor. Moreover $\alpha$,
$\beta$, $\gamma$ and $\varphi$ in Eq. (\ref{perturbationscosmo})
denote the scalar type order variables of the cosmological
perturbations, while the tensor $C_{\alpha \beta}$ stands for the
traceless and transverse tensor perturbation. The dynamical
evolution of the scalar type perturbation quantified by the
variable $\Phi=\varphi_{\delta \phi}$, and which is directly
related to the scalar type gauge invariant quantity $\delta
\phi_{\varphi}=-\frac{\dot{\phi}}{H}\varphi_{\delta \phi}$, is
determined by the differential equation that follows,
\begin{equation}\label{bigdiffeeqannew}
\frac{\left(H+\frac{\dot{F}_R}{2F_R}\right)^2}{a^3\left(\omega\dot{\phi}^2+3\frac{(\dot{F}_R)^2}{2F_R}
\right)}\frac{d}{d t} \left
(\frac{a^3\left(\omega\dot{\phi}^2+3\frac{(\dot{F}_R)^2}{2F_R}\right)}{\left(H+\frac{\dot{F}_R}{2F_R}\right)^2}
\dot{\Phi}\right )=\frac{\Delta}{a^2}\Phi\, ,
\end{equation}
with $\Delta$ being the Laplacian of the spatial section of the
FRW metric. It is apparent from the right hand side of the above
evolution equation that the wave speed of the scalar perturbations
is equal to unity, thus no ghost instabilities are expected in the
cosmological perturbations. But why does the wave speed determines
actually whether on not ghost instabilities occur in the first
place? Basically, by looking at Eq. (\ref{bigdiffeeqannew}), the
wave speed of the perturbation $\Phi=\varphi_{\delta \phi}$ is
equal to unity, but apart from the perturbation $\Phi$ for a
general $f(R,\phi)$ theory, there is another scalar perturbation,
denoted as $\Psi$, which is defined as follows,
\begin{equation}\label{psidef}
\Psi=\varphi_{\chi}+\frac{\dot{F}_R}{2F_R}\frac{\delta
F_{\chi}}{\dot{F}}\, ,
\end{equation}
the evolution of which is determined by the differential equation
that follows,
\begin{align}\label{bigevolutioneqn}
& \frac{\omega
\dot{\phi}^2+3\frac{(\dot{F}_R)^2}{2F}}{\left(H\frac{\dot{F}_R}{2F_R}
\right)(F_R)}\times \\ \notag &
\frac{d}{dt}\Big{(}\frac{\left(H+\frac{\dot{F}_R}{2F_R}
\right)^2}{a\left(\omega
\dot{\phi}^2+3\frac{(\dot{F}_R)^2}{2F_R}\right)}\dot{\mathcal{S}}\Big{)}=\frac{\Delta}{a^2}\Psi\,
,
\end{align}
with the quantity $\mathcal{S}$ being defined below,
\begin{equation}\label{mathcals}
\mathcal{S}=\frac{a(F_R)}{H+\frac{\dot{F}_R}{2F_R}}\Psi\, .
\end{equation}
Hence, by looking both Eqs. (\ref{bigdiffeeqannew}) and
(\ref{bigevolutioneqn}), which are essentially wave equations, the
propagation wave speed is equal to unity, so the wave speed of the
propagation of both the fluctuating field and of the perturbed
metric is equal to unity. In fact, since the FRW metric we use has
a flat spatial part, the sound wave speed defined as
$c_s^2=\frac{\dot{p}}{\dot{\rho}}$ is identical to the wave speed,
which is equal to unity. In order to make this more apparent, let
us bring both Eqs. (\ref{bigdiffeeqannew}) and
(\ref{bigevolutioneqn}) to the Mukhanov-Sasaki form. We define
$\bar{z}=c_Az$, with $c_A=1$ in our case, and $z$ is,
\begin{equation}\label{zetadef}
z=\frac{a\dot{\phi}}{H}\sqrt{\frac{E}{F_R}}\, ,
\end{equation}
while $E$ is,
\begin{equation}
\centering \label{E1}
E=\frac{F_R}{\kappa^2}\left(1+\frac{3(\dot{F}_R)^2}{2\kappa^2\dot\phi^2F_R}\right)\,
.
\end{equation}
In addition, we introduce $\bar{v}=z\Phi$ and also
$u=\frac{a}{\kappa^2H}\frac{1}{\bar{z}}\Psi$, hence the wave
equations appearing in Eqs. (\ref{bigdiffeeqannew}) and
(\ref{bigevolutioneqn}) respectively, can be brought in the
well-known Mukhanov-Sasaki forms,
\begin{equation}\label{msform}
\bar{v}''-\left(c_A^2\Delta+\frac{z''}{z}\right)\bar{v}=0\, ,
\end{equation}
\begin{equation}\label{msform1}
u''-\left(c_A^2\Delta+\frac{(1/\bar{z})''}{1/\bar{z}}\right)u=0\,
,
\end{equation}
where in our case $c_A=1$. Thus the wave speed of the fluctuating
fluid and of the perturbed metric is equal to unity, hence no
ghost instabilities occur in the theoretical framework we used.

\section{Concluding Remarks}

In this work we investigated the phenomenology of an $F(R)$
gravity model in the presence of a primordial light axion scalar
field with broken $U(1)$ Peccei-Quinn symmetry during the
inflationary era. Due to the fact that the axion scalar is frozen
to its primordial vacuum expectation values, it has a highly
subdominant role during the inflationary era, which is
predominantly controlled by the $R^2$ term of the $F(R)$ gravity.
The resulting effective Lagrangian during inflation, has also a
rescaled Einstein-Hilbert $R$ term at leading order during
inflation. As we demonstrated by performing an explicit
calculation, the resulting inflationary phenomenology is identical
with that corresponding to the $R^2$ model, thus surprisingly the
rescaled $R$ term does not affect the dynamical evolution during
inflation, at least at leading order. However, if the axion was
dynamically evolving during inflation, this result would not be
valid. Actually, the rescaled Einstein-Hilbert gravity can
significantly affect the inflationary phenomenology of a canonical
scalar field as we will show in \cite{workinprogress}. After the
inflationary era, and specifically when the axion mass $m_a$
becomes of the same order as the Hubble rate, and for all cosmic
times for which $m_a\succeq H$, the axion field starts to
oscillate and its energy density evolves as $\rho_a\sim a^{-3}$,
thus behaves as a dark matter perfect fluid. Accordingly we
examined the dark energy phenomenology of the $F(R)$ gravity model
in the presence of the axion dark matter perfect fluid and of the
radiation perfect fluid. After expressing the Friedmann equation
as a function of the redshift and of a suitable statefinder
function, we solved numerically the resulting equation, using
suitable physically motivated initial conditions corresponding to
the late stages of the matter domination era. For the numerical
analysis we developed a PYTHON 3 numerical code which is freely
available here \cite{code}, along with a pedagogical description
of the $F(R)$ gravity dark energy phenomenology. The resulting
picture is quite interesting phenomenologically, since three novel
features for $F(R)$ gravity appear in the theory. Firstly, the
Universe experiences an early dark energy era, starting at $z\sim
2.5$ and ending at $z\sim 1.5$, followed by a decelerating epoch
which lasts until $z\sim 0.5$, and for the approximately remaining
5 billion years, from $z\sim 0.5$ to $z=0$, the Universe
accelerates again. We found that the model is also viable and
compatible with the 2018 Planck constraints on the cosmological
parameters, at least when the dark energy density parameter and
the dark energy EoS parameters are considered. Also the model
mimics the $\Lambda$CDM model significantly at the last stages of
the evolution prior to present day redshift $z=0$, and at least
qualitatively. Secondly, the model is free from dark energy
oscillations, which are known to plague the $F(R)$ gravity dark
energy phenomenology. Thus our result indicates that the dark
energy oscillations might be a model dependent feature of $F(R)$
gravity dark energy phenomenology. Thirdly, the model also
complies with the $z\sim 2.34$ observation of
\cite{Delubac:2014aqe}. Finally, the model has also another
interesting characteristic occurring at $z\sim 2$, where the
Hubble rate has a local minimum, followed by a local maximum,
after which evolves normally up to $z=0$. This however is another
model dependent feature. An interesting phenomenological concept
which emerged from the effective Lagrangian of the present $F(R)$
gravity model at early times, is the appearance of a rescaled
Einstein-Hilbert term $\alpha R$. Although for the present work,
this term played a subdominant term during inflation, due to the
presence of the $R^2$ term, in conjunction with the fact that the
axion scalar was frozen in its vacuum expectation value, it would
be interesting to consider the same $F(R)$ gravity model without
the $R^2$ term, and in the presence of a dynamically evolving
canonical scalar field. As we shall show in \cite{workinprogress},
the results can be quite intriguing in some cases of interest.

\section*{Appendix: The Python Code for $F(R)$ Gravity Dark Energy Phenomenology}

\begin{verbatim}


import numpy as np

import sympy as sp

import matplotlib.pyplot as plt

from scipy.integrate import solve_ivp

 from scipy.integrate import odeint

 from scipy.misc import derivative

#Definition of Hubble rate at present time based on CMB Planck
#2018 data. It is in eV

H0p =
(67.4)*(5.067*10**5*10**13*10**9)/(1.592*1.5637*10**24*10**38)

#Definition of Hubble rate at present time based on Cepheids data.
It is in eV

H0c = (74)*(5.067*10**5*10**13*10**9)/(1.592*1.5637*10**24*10**38)

#Mass scale (eV)

ms = 1.87101*10**(-67)

#Cosmological constant (eV)

Lambda_s= 7.93*1.5*10**(-67)

# the parameter $\chi$, radiation over dark matter present day
energy densities.

chi=3.1*10**(-4)

#The initial redshift value of the redefined redshift z'=0, which
corresponds to the final value of the original redshift z=10

zfin = 0

#initial conditions #Notice that we used the redefined redshift
z'=10-z, so the value zfin=0 of the redefined redshift corresponds
# to zfin=10 of the initial redshift parameter.

r = Lambda_s/(3*ms)*(1 + (11-zfin)/1000)

phi= -Lambda_s/(3*ms)*1/1000

#Model parameters

M = 3.04375*10**22 # in (eV)

#lambda_s=0.01 #dimensionless lambda_s=0.9999 #dimensionless

#t=z from now on in order for the solve_ivp function to work. #Now
you can use fR in your programm

def f(t,y):
    yH=y[0]
    Y=y[1]
    #first definition of the curvature for the numerical integration.
This must be re-defined after the numerical integration
    #in order to be expressed in terms of the new solution
    a=1/(11-t)
    rs=(chi)*a**(-4)
    R=3*ms*((1/a)*Y + 4*yH + a**(-3))
    #The fR and fRR are taken from above
    f=R**2/M**2+lambda_s*R*np.exp(-Lambda_s*5.1/R)+Lambda_s*5.1*lambda_s-(Lambda_s/0.2)*(R/(ms))**0.1
    fR=5.1*Lambda_s*lambda_s*np.exp(-5.1*Lambda_s/R)/R -
0.5*Lambda_s*(R/ms)**0.1/R + lambda_s*np.exp(-5.1*Lambda_s/R) +
2*R/M**2
    fRR=26.01*Lambda_s**2*lambda_s*np.exp(-5.1*Lambda_s/R)/R**3 +
0.45*Lambda_s*(R/ms)**0.1/R**2 + 2/M**2
    #The fR and fRR are taken from above
    J1=a*(-3*fRR - (1/(yH + a**(-3) + rs))*(-fR/(6*ms)))
    J2=(a**2)*(1/(yH + a**(-3) + rs))*(1 - fR)/(3*ms)
    J3=-3*a**(-1)*fRR - ((a**2)*(-fR*(a**(-3) + 2*rs )- f/(3*ms))/(yH
+ a**(-3) + rs))*1/(6*ms)
    #we denote the derivative of y_H with respect to the redefined
redshift z' as dyHdzr. This is one of the two things
    #that the f function will give us
    dyHdz=Y
    dYdz=-(1/fRR)*(-J1*Y+J2*yH+J3)
    return np.array([dyHdz,dYdz])

# At this point we set the redefined redshift initial and final
#values: from z=0 to z=10. t_span=np.array([0,10])

#And at this point we choose all the intermediate redshift points
#from z=0 to z=10, using 1000 points. In principle one #can choose
#more intermediate points or even less, it is up to the reader.

time_interval=np.linspace(t_span[0],t_span[1],1000)

# Now we set the initial condition. Since y the outcome of the
#solve_ivp that will follow, is an array, with y[0] being the
#function y_H and y[1] being the derivative yH'[z], the initial
#condition is an array fixing the initial value of yH to be r
#defined previously, and that of yH'[z] to be phi, also defined
#previously, below the definition of r.

y0=np.array([r,phi])

# Now we can the solve_ivp function of the scipy module:

soly=solve_ivp(f,t_span,y0,method='LSODA',t_eval=time_interval)

# The solution yH is y[0], so we assign the name yH to the
#solution y of the solve_ivp function which we named ''sol'' #The
#same with yH'[z] # which we call dotyH

yH=soly.y[0] Y=soly.y[1]

#Now since the above are basically arrays, we switch the redshift
#to the initial definition by using the very own definition # of
#the initial redshift-in our case recall we had to identify z=t in
#order for the solve_ivp function to work.

t=10-soly.t

#Now the definition of the Hubble rate of the present model
#defined as a function of the function yH and of the redshift.
#The outcome is an array, to be #plotted as a function of the
#initially defined redshift which in terms of the programming
#dynamical variable t is t=10-soly.t # look up above this comment.

def a(x):
    return 1/(11-x)

# Now we need to define some functions of the scale factor and of
#the functions yH and dotyH. These must be evaluated for the
#initial time interval, not the redefined t, because the functions
#yH and dotyH were calculated in exactly this way. In the #plots
#we shall make use of the initial redshift values, but in the
#plots we basically connect values of arrays, #a list plot
#basically, so t=10-soly.t is just an array of values. However,
#the functions that will be built depending on yH and #dotyH MUST
#be evaluated using the initial time interval, in our case the
#linspace time_interval

rs=(chi)*a(time_interval)**(-4)

H= np.sqrt(ms*(yH + a(time_interval)**(-3) + rs))

# The derivative of the Hubble rate

dHdz = (ms*(-3*(a(time_interval))**(-2) -
0.00124*a(time_interval)**(-3) + Y))/(2*np.sqrt(ms*(yH +
a(time_interval)**(-3) + rs)))

Omega_Lambda = 0.6847

 Omega_R = 8*10**(-5)

Omega_M = 0.3153 - Omega_R

#Dark energy density parameter

Omega_DE=yH/(yH+ a(time_interval)**(-3) + rs)

#Hubble Rate of LCDM

HLCDM=np.sqrt(H0p**2*(Omega_Lambda+ Omega_M*
a(time_interval)**(-3) + Omega_R*a(time_interval)**(-4)))

#The derivative of the Hubble rate

dHLdz = -(H0p**2*(3*a(time_interval)**(-2)*Omega_M +
4*a(time_interval)**(-3)*Omega_R))/(2*np.sqrt(H0p**2*((a(time_interval))**(-3)*Omega_M+
(a(time_interval))**(-4)*Omega_R + Omega_Lambda)))

#Total Effective Equation of State parameter (Total EoS Parameter)

# For the $F(R)$ gravity model

weff=-1 - 2/3*a(time_interval)**(-1)*dHdz/H

#For the LCDM model

wefflcdm=-1 -2/3*a(time_interval)**(-1)*dHLdz/HLCDM;

#Dark Energy Equation of State Parameter:

wG=-1 -(1/3)*a(time_interval)**(-1)*Y/yH

#DECELERATION Parameter for the F(Rs) Gravity Model

q= -(a(time_interval)**(-1)/H)*dHdz - 1

#DECELERATION for the LCDM Model

qlcdm=-(a(time_interval)**(-1)/HLCDM)*dHLdz - 1

# Now the plots and the results of the numerical analysis # The
#F(R)  model corresponds to blue color, and dashed-dot ''-.''
#curve #The LCDM model corresponds to red color and double dashed
#'--' curve.

#plt.style.use('Solarize_Light2')

plt.plot(time_interval,yH,marker='.',color='g',linestyle='-.',label='Programming
Method')

plt.plot(t,yH,marker='.',color='b',linestyle='--',label='Correct
Redshift Method')

plt.title('The statefinder $y_H$ vs Redshift' )

plt.xlabel('z') plt.ylabel('$y_H$')

plt.tight_layout()

plt.grid(True)

plt.legend()

#plt.savefig('plot1.eps')

#plt.savefig('plot1.jpg')

plt.show()

plt.plot(time_interval,-Y,marker='.',color='g',linestyle='-.',label='Programming
Method')

plt.plot(t,-Y,marker='.',color='b',linestyle='--',label='Correct
Redshift Method')

plt.title('The statefinder derivative $y\'_H$ vs Redshift' )

plt.xlabel('z') plt.ylabel('$y\'_H$')

plt.tight_layout()

plt.grid(True)

plt.legend()

#plt.savefig('dyHp.jpg')

plt.show()

plt.plot(t,H/H0p,marker='.',color='b',linestyle='-.',label='$f(R)$')

plt.plot(t,HLCDM/H0p,marker='.',color='r',linestyle='--',label='$\Lambda$CDM')

plt.title(' $H/H_0$ vs Redshift' )

plt.xlabel('z') plt.ylabel('Hubble Rates')

plt.tight_layout()

plt.grid(True)

plt.legend()

#plt.savefig('Hubblep.jpg')

plt.show()

Ht=1.996*10**(-33)/H0p*np.ones(1000)

plt.plot(t,H/H0p,marker='.',color='b',linestyle='-.',label='$f(R)$')

plt.plot(t,HLCDM/H0p,marker='.',color='r',linestyle='--',label='$\Lambda$CDM')

plt.plot(t,Ht,marker='.',color='g',label='$H_t=1.996 10^{-33}eV$')

plt.title(' $H/H_0$ vs Redshift Close Up' )

plt.xlabel('z') plt.ylabel('Hubble rates over $H_0$: H/H_0')

plt.tight_layout()

plt.grid(True)

plt.legend()

#plt.savefig('Hubblepc.jpg')

plt.ylim(1, 2) plt.xlim(1.5,3)

plt.show()

plt.plot(t,Omega_DE,marker='.',color='b',linestyle='-.',label='$\Omega_{DE}(z)$-z
for the f(R) Model')


plt.title('$\Omega_{DE}(z)$ vs Redshift' )


plt.xlabel('z') plt.ylabel('$\Omega_{DE}(z)$')

plt.tight_layout()

plt.grid(True)

plt.legend()

#plt.savefig('Omdep.jpg')

plt.show()

plt.plot(t,weff,marker='.',color='b',linestyle='-.',label='$\omega_{eff}(z)$-z
for the f(R) Model')

plt.plot(t,wefflcdm,marker='.',color='r',linestyle='--',label='$\omega_{eff}(z)$-z
for the $\Lambda$CDM Model')

plt.title('Total EoS vs Redshift')

plt.xlabel('z') plt.ylabel('EoS Parameters')

plt.tight_layout()

plt.grid(True)

plt.legend()

#plt.savefig('plot2.eps')

#plt.savefig('totaleosp.jpg')

plt.show()

plt.plot(t,wG,marker='.',color='b',linestyle='-.',label='$\omega_{DE}(z)$-z
for the f(R) Model')

plt.title('Dark Energy EoS vs Redshift')

plt.xlabel('z') plt.ylabel('EoS Parameters')

plt.tight_layout()

plt.grid(True)

plt.legend()

#plt.savefig('odep.jpg')

plt.show()

plt.plot(t,q,marker='.',color='b',linestyle='-.',label='$q$-z for
the $f(R)$ Model')

plt.plot(t,qlcdm,marker='.',color='r',linestyle='-.',label='$q$-z
for the $\Lambda$CDM Model')

plt.title('Deceleration Parameter vs Redshift')

plt.xlabel('z') plt.ylabel('q')

plt.tight_layout()

plt.grid(True)

plt.legend()

#plt.savefig('plot3.eps')

#plt.savefig('decelerationp.jpg')

plt.show()

print("The dark energy density parameter $\Omega_DE(0)$ for the
$f(R)$ Model at present time is:"+" "+str(Omega_DE[999]))
 print()

print()

 print("The dark energy EoS parameter for the $f(R)$ Model
at present time is:"+" "+str(wG[999]))

 print()

  print()

   print("The
Hubble rate for the $f(R)$ Model at present time is:"+"
"+str(H[999]))

print()

print()

 print("The fraction of the Hubble rate for the $f(R)$
Model and of the $\Lmabda$CDM model at present time is:"+"
"+str(H[999]/H0p))

print() print()

print("The deceleration parameter for the $f(R)$ Model at present
time is:"+" "+str(q[999]))

 print()

\end{verbatim}


\begin{thebibliography}{99}




\bibitem{Riess:1998cb}
  A.~G.~Riess {\it et al.} [Supernova Search Team],
  Astron.\ J.\  {\bf 116} (1998) 1009
  [astro-ph/9805201].




\bibitem{Bertone:2004pz}
  G.~Bertone, D.~Hooper and J.~Silk,
  Phys.\ Rept.\  {\bf 405} (2005) 279
  doi:10.1016/j.physrep.2004.08.031
  [hep-ph/0404175].


\bibitem{Bergstrom:2000pn}
  L.~Bergstrom,
  Rept.\ Prog.\ Phys.\  {\bf 63} (2000) 793
  doi:10.1088/0034-4885/63/5/2r3
  [hep-ph/0002126].




\bibitem{Mambrini:2015sia}
  Y.~Mambrini, S.~Profumo and F.~S.~Queiroz,
  Phys.\ Lett.\ B {\bf 760} (2016) 807
  [arXiv:1508.06635 [hep-ph]].

\bibitem{Profumo:2013yn}
  S.~Profumo,
  arXiv:1301.0952 [hep-ph].




\bibitem{Hooper:2007qk}
  D.~Hooper and S.~Profumo,
  Phys.\ Rept.\  {\bf 453} (2007) 29
  [hep-ph/0701197].



\bibitem{Oikonomou:2006mh}
V.~K.~Oikonomou, J.~D.~Vergados and C.~C.~Moustakidis,
Nucl.\ Phys.\ B {\bf 773} (2007) 19
[hep-ph/0612293].




\bibitem{Nojiri:2003ft}
S.~Nojiri and S.~D.~Odintsov,
Phys.\ Rev.\ D {\bf 68} (2003) 123512
doi:10.1103/PhysRevD.68.123512 [hep-th/0307288].


\bibitem{Nojiri:2007as}
S.~Nojiri and S.~D.~Odintsov,
Phys.\ Lett.\ B {\bf 657} (2007) 238
doi:10.1016/j.physletb.2007.10.027 [arXiv:0707.1941 [hep-th]].

\bibitem{Nojiri:2007cq}
S.~Nojiri and S.~D.~Odintsov,
Phys.\ Rev.\ D {\bf 77} (2008) 026007
doi:10.1103/PhysRevD.77.026007 [arXiv:0710.1738 [hep-th]].

\bibitem{Cognola:2007zu}
G.~Cognola, E.~Elizalde, S.~Nojiri, S.~D.~Odintsov, L.~Sebastiani
and S.~Zerbini,
Phys.\ Rev.\ D {\bf 77} (2008) 046009
doi:10.1103/PhysRevD.77.046009 [arXiv:0712.4017 [hep-th]].

\bibitem{Nojiri:2006gh}
S.~Nojiri and S.~D.~Odintsov,
Phys.\ Rev.\ D {\bf 74} (2006) 086005
doi:10.1103/PhysRevD.74.086005 [hep-th/0608008].

\bibitem{Appleby:2007vb}
S.~A.~Appleby and R.~A.~Battye,
Phys.\ Lett.\ B {\bf 654} (2007) 7
doi:10.1016/j.physletb.2007.08.037 [arXiv:0705.3199 [astro-ph]].



\bibitem{Elizalde:2010ts}
E.~Elizalde, S.~Nojiri, S.~D.~Odintsov, L.~Sebastiani and
S.~Zerbini,
Phys.\ Rev.\ D {\bf 83} (2011) 086006
doi:10.1103/PhysRevD.83.086006 [arXiv:1012.2280 [hep-th]].

\bibitem{Odintsov:2020nwm}
  S.~D.~Odintsov and V.~K.~Oikonomou,
  arXiv:2001.06830 [gr-qc].

\bibitem{Sa:2020fvn}
P.~M.~S\'a,
Phys. Rev. D \textbf{102} (2020) no.10, 103519
doi:10.1103/PhysRevD.102.103519 [arXiv:2007.07109 [gr-qc]].



\bibitem{reviews1}
 S.~Nojiri, S.~D.~Odintsov and V.~K.~Oikonomou,
  Phys.\ Rept.\  {\bf 692} (2017) 1
  [arXiv:1705.11098 [gr-qc]].

\bibitem{reviews2}


 S. Capozziello, M. De Laurentis,
   Phys.\ Rept.\  {\bf 509}, 167 (2011);\\
 V.~Faraoni and S.~Capozziello,
  Fundam.\ Theor.\ Phys.\  {\bf 170} (2010).



\bibitem{reviews3}
S. Nojiri, S.D. Odintsov,
  eConf {\bf C0602061}, 06 (2006)
  [Int.\ J.\ Geom.\ Meth.\ Mod.\ Phys.\  {\bf 4}, 115 (2007)].


   \bibitem{reviews4}

S. Nojiri, S.D. Odintsov,
   Phys.\ Rept.\  {\bf 505}, 59 (2011);




\bibitem{reviews5}

A.~de la Cruz-Dombriz and D.~Saez-Gomez,
  Entropy {\bf 14} (2012) 1717
  [arXiv:1207.2663 [gr-qc]].

\bibitem{reviews6}

G.~J.~Olmo,
  Int.\ J.\ Mod.\ Phys.\ D {\bf 20} (2011) 413
  [arXiv:1101.3864 [gr-qc]].


\bibitem{Preskill:1982cy}
  J.~Preskill, M.~B.~Wise and F.~Wilczek,
  Phys.\ Lett.\  {\bf 120B} (1983) 127.
  doi:10.1016/0370-2693(83)90637-8


\bibitem{Abbott:1982af}
  L.~F.~Abbott and P.~Sikivie,
  Phys.\ Lett.\  {\bf 120B} (1983) 133.
  doi:10.1016/0370-2693(83)90638-X


\bibitem{Dine:1982ah}
  M.~Dine and W.~Fischler,
  Phys.\ Lett.\  {\bf 120B} (1983) 137.
  doi:10.1016/0370-2693(83)90639-1







\bibitem{Marsh:2015xka}
  D.~J.~E.~Marsh,
  Phys.\ Rept.\  {\bf 643} (2016) 1
  [arXiv:1510.07633 [astro-ph.CO]].






\bibitem{Sikivie:2006ni}
  P.~Sikivie,
  Lect.\ Notes Phys.\  {\bf 741} (2008) 19
  [astro-ph/0610440].



\bibitem{Raffelt:2006cw}
  G.~G.~Raffelt,
  Lect.\ Notes Phys.\  {\bf 741} (2008) 51
  [hep-ph/0611350].


\bibitem{Linde:1991km}
  A.~D.~Linde,
  Phys.\ Lett.\ B {\bf 259} (1991) 38.



\bibitem{Co:2019jts}
R.~T.~Co, L.~J.~Hall and K.~Harigaya,
Phys. Rev. Lett. \textbf{124} (2020) no.25, 251802
doi:10.1103/PhysRevLett.124.251802 [arXiv:1910.14152 [hep-ph]].


\bibitem{Marsh:2017yvc}
  M.~C.~D.~Marsh, H.~R.~Russell, A.~C.~Fabian, B.~P.~McNamara, P.~Nulsen and C.~S.~Reynolds,
  JCAP {\bf 1712} (2017) no.12,  036
  [arXiv:1703.07354 [hep-ph]].







\bibitem{Odintsov:2019mlf}
  S.~D.~Odintsov and V.~K.~Oikonomou,
  Phys.\ Rev.\ D {\bf 99} (2019) no.6,  064049
  [arXiv:1901.05363 [gr-qc]].




\bibitem{Nojiri:2019nar}
  S.~Nojiri, S.~D.~Odintsov, V.~K.~Oikonomou and A.~A.~Popov,
  Phys.\ Rev.\ D {\bf 100} (2019) no.8,  084009
  [arXiv:1909.01324 [gr-qc]].


\bibitem{Nojiri:2019riz}
S.~Nojiri, S.~D.~Odintsov and V.~K.~Oikonomou,
Annals Phys. \textbf{418} (2020), 168186
doi:10.1016/j.aop.2020.168186 [arXiv:1907.01625 [gr-qc]].

\bibitem{Odintsov:2019evb}
  S.~D.~Odintsov and V.~K.~Oikonomou,
  Phys.\ Rev.\ D {\bf 99} (2019) no.10,  104070
  [arXiv:1905.03496 [gr-qc]].






\bibitem{Cicoli:2019ulk}
  M.~Cicoli, V.~Guidetti and F.~G.~Pedro,
  arXiv:1903.01497 [hep-th].

\bibitem{Fukunaga:2019unq}
  H.~Fukunaga, N.~Kitajima and Y.~Urakawa,
  arXiv:1903.02119 [astro-ph.CO].


\bibitem{Caputo:2019joi}
  A.~Caputo,
  arXiv:1902.02666 [hep-ph].





\bibitem{maxim}

A.S.Sakharov and M.Yu.Khlopov, 
 Yadernaya Fizika (1994) V. 57, PP. 514-
516. ( Phys.Atom.Nucl. (1994) V. 57, PP. 485-487); A.S.Sakharov,
D.D.Sokoloff and M.Yu.Khlopov, 
 Yadernaya Fizika (1996) V. 59, PP. 1050-1055.
(Phys.Atom.Nucl. (1996) V. 59, PP. 1005-1010); M .Yu.Khlopov,
A.S.Sakharov and D.D.Sokoloff,
 Nucl.Phys. B (Proc. Suppl.) (1999) V. 72, 105-109.







\bibitem{Chang:2018rso}
  J.~H.~Chang, R.~Essig and S.~D.~McDermott,
  JHEP {\bf 1809} (2018) 051
  [arXiv:1803.00993 [hep-ph]].

Chang:2018rso,Irastorza:2018dyq,

\bibitem{Irastorza:2018dyq}
  I.~G.~Irastorza and J.~Redondo,
  Prog.\ Part.\ Nucl.\ Phys.\  {\bf 102} (2018) 89
  [arXiv:1801.08127 [hep-ph]].


\bibitem{Anastassopoulos:2017ftl}
  V.~Anastassopoulos {\it et al.} [CAST Collaboration],
  Nature Phys.\  {\bf 13} (2017) 584
  [arXiv:1705.02290 [hep-ex]].








\bibitem{Sikivie:2014lha}
  P.~Sikivie,
  Phys.\ Rev.\ Lett.\  {\bf 113} (2014) no.20,  201301
  [arXiv:1409.2806 [hep-ph]].






\bibitem{Sikivie:2010bq}
  P.~Sikivie,
  Phys.\ Lett.\ B {\bf 695} (2011) 22
  [arXiv:1003.2426 [astro-ph.GA]].




\bibitem{Sikivie:2009qn}
  P.~Sikivie and Q.~Yang,
  Phys.\ Rev.\ Lett.\  {\bf 103} (2009) 111301
  [arXiv:0901.1106 [hep-ph]].



\bibitem{Caputo:2019tms}
  A.~Caputo, L.~Sberna, M.~Frias, D.~Blas, P.~Pani, L.~Shao and W.~Yan,
  Phys.\ Rev.\ D {\bf 100} (2019) no.6,  063515
  [arXiv:1902.02695 [astro-ph.CO]].




\bibitem{Masaki:2019ggg}
  E.~Masaki, A.~Aoki and J.~Soda,
  arXiv:1909.11470 [hep-ph].


\bibitem{Soda:2017sce}
  J.~Soda and D.~Yoshida,
  Galaxies {\bf 5} (2017) no.4,  96.


\bibitem{Soda:2017dsu}
  J.~Soda and Y.~Urakawa,
  Eur.\ Phys.\ J.\ C {\bf 78} (2018) no.9,  779
  [arXiv:1710.00305 [astro-ph.CO]].




\bibitem{Aoki:2017ehb}
  A.~Aoki and J.~Soda,
  Phys.\ Rev.\ D {\bf 96} (2017) no.2,  023534
  [arXiv:1703.03589 [astro-ph.CO]].










\bibitem{Masaki:2017aea}
  E.~Masaki, A.~Aoki and J.~Soda,
  Phys.\ Rev.\ D {\bf 96} (2017) no.4,  043519
  [arXiv:1702.08843 [astro-ph.CO]].


\bibitem{Aoki:2016kwl}
  A.~Aoki and J.~Soda,
  Int.\ J.\ Mod.\ Phys.\ D {\bf 26} (2016) no.07,  1750063
  [arXiv:1608.05933 [astro-ph.CO]].


\bibitem{Obata:2016xcr}
  I.~Obata and J.~Soda,
  Phys.\ Rev.\ D {\bf 94} (2016) no.4,  044062
  [arXiv:1607.01847 [astro-ph.CO]].


\bibitem{Aoki:2016mtn}
  A.~Aoki and J.~Soda,
  Phys.\ Rev.\ D {\bf 93} (2016) no.8,  083503
  [arXiv:1601.03904 [hep-ph]].



\bibitem{Ikeda:2019fvj}
  T.~Ikeda, R.~Brito and V.~Cardoso,
  Phys.\ Rev.\ Lett.\  {\bf 122} (2019) no.8,  081101
  [arXiv:1811.04950 [gr-qc]].



\bibitem{Arvanitaki:2019rax}
  A.~Arvanitaki, S.~Dimopoulos, M.~Galanis, L.~Lehner, J.~O.~Thompson and K.~Van Tilburg,
  arXiv:1909.11665 [astro-ph.CO].



\bibitem{Arvanitaki:2016qwi}
  A.~Arvanitaki, M.~Baryakhtar, S.~Dimopoulos, S.~Dubovsky and R.~Lasenby,
  Phys.\ Rev.\ D {\bf 95} (2017) no.4,  043001
  [arXiv:1604.03958 [hep-ph]].

\bibitem{Arvanitaki:2014wva}
  A.~Arvanitaki, M.~Baryakhtar and X.~Huang,
  Phys.\ Rev.\ D {\bf 91} (2015) no.8,  084011
  [arXiv:1411.2263 [hep-ph]].


\bibitem{Arvanitaki:2014dfa}
  A.~Arvanitaki and A.~A.~Geraci,
  Phys.\ Rev.\ Lett.\  {\bf 113} (2014) no.16,  161801
  [arXiv:1403.1290 [hep-ph]].




\bibitem{Sen:2018cjt}
  S.~Sen,
  Phys.\ Rev.\ D {\bf 98} (2018) no.10,  103012
  [arXiv:1805.06471 [hep-ph]].









\bibitem{Cardoso:2018tly}
  V.~Cardoso, S.~J.~C.~Dias, G.~S.~Hartnett, M.~Middleton, P.~Pani and J.~E.~Santos,
  JCAP {\bf 1803} (2018) 043
  [arXiv:1801.01420 [gr-qc]].


\bibitem{Rosa:2017ury}
  J.~G.~Rosa and T.~W.~Kephart,
  Phys.\ Rev.\ Lett.\  {\bf 120} (2018) no.23,  231102
  [arXiv:1709.06581 [gr-qc]].


\bibitem{Yoshino:2013ofa}
  H.~Yoshino and H.~Kodama,
  PTEP {\bf 2014} (2014) 043E02
  [arXiv:1312.2326 [gr-qc]].



\bibitem{Machado:2019xuc}
  C.~S.~Machado, W.~Ratzinger, P.~Schwaller and B.~A.~Stefanek,
  arXiv:1912.01007 [hep-ph].




\bibitem{Korochkin:2019qpe}
  A.~Korochkin, A.~Neronov and D.~Semikoz,
  arXiv:1911.13291 [hep-ph].


\bibitem{Chou:2019enw}
  A.~S.~Chou,
  Astrophys.\ Space Sci.\ Proc.\  {\bf 56} (2019) 41.


\bibitem{Chang:2019tvx}
  C.~F.~Chang and Y.~Cui,
  arXiv:1911.11885 [hep-ph].








\bibitem{Crisosto:2019fcj}
  N.~Crisosto, G.~Rybka, P.~Sikivie, N.~S.~Sullivan, D.~B.~Tanner and J.~Yang,
  arXiv:1911.05772 [astro-ph.CO].


\bibitem{Choi:2019jwx}
  K.~Choi, H.~Seong and S.~Yun,
  arXiv:1911.00532 [hep-ph].








\bibitem{Kavic:2019cgk}
  M.~Kavic, S.~L.~Liebling, M.~Lippert and J.~H.~Simonetti,
  arXiv:1910.06977 [astro-ph.HE].


\bibitem{Blas:2019qqp}
  D.~Blas, A.~Caputo, M.~M.~Ivanov and L.~Sberna,
  arXiv:1910.06128 [hep-ph].




\bibitem{Guerra:2019srj}
  D.~Guerra, C.~F.~B.~Macedo and P.~Pani,
  JCAP {\bf 1909} (2019) no.09,  061
  [arXiv:1909.05515 [gr-qc]].




\bibitem{Tenkanen:2019xzn}
  T.~Tenkanen and L.~Visinelli,
  JCAP {\bf 1908} (2019) 033
  [arXiv:1906.11837 [astro-ph.CO]].




\bibitem{Huang:2019rmc}
  G.~Y.~Huang and S.~Zhou,
  Phys.\ Rev.\ D {\bf 100} (2019) no.3,  035010
  [arXiv:1905.00367 [hep-ph]].




\bibitem{Croon:2019iuh}
  D.~Croon, R.~Houtz and V.~Sanz,
  JHEP {\bf 1907} (2019) 146
  [arXiv:1904.10967 [hep-ph]].


\bibitem{Day:2019bbh}
  F.~V.~Day and J.~I.~McDonald,
  JCAP {\bf 1910} (2019) no.10,  051
  [arXiv:1904.08341 [hep-ph]].


\bibitem{Odintsov:2020iui}
S.~D.~Odintsov and V.~K.~Oikonomou,
EPL \textbf{129} (2020) no.4, 40001
doi:10.1209/0295-5075/129/40001 [arXiv:2003.06671 [gr-qc]].



\bibitem{Nojiri:2020pqr}
S.~Nojiri, S.~D.~Odintsov, V.~K.~Oikonomou and A.~A.~Popov,
Phys. Dark Univ. \textbf{28} (2020), 100514
doi:10.1016/j.dark.2020.100514 [arXiv:2002.10402 [gr-qc]].






\bibitem{Akrami:2018odb}
  Y.~Akrami {\it et al.} [Planck Collaboration],
  arXiv:1807.06211 [astro-ph.CO].


\bibitem{code}

https://github.com/VOikonomou?tab=projects



\bibitem{Aghanim:2018eyx}
  N.~Aghanim {\it et al.} [Planck Collaboration],
  arXiv:1807.06209 [astro-ph.CO].





\bibitem{Doran:2006kp}
M.~Doran and G.~Robbers,
JCAP {\bf 0606} (2006) 026 doi:10.1088/1475-7516/2006/06/026
[astro-ph/0601544].

\bibitem{Bhattacharyya:2019lvg}
A.~Bhattacharyya and S.~Pal,
arXiv:1907.10946 [astro-ph.CO].

\bibitem{Sakstein:2019fmf}
J.~Sakstein and M.~Trodden,
arXiv:1911.11760 [astro-ph.CO].


\bibitem{Tian:2019enx}
  S.~X.~Tian,
  arXiv:1912.13208 [astro-ph.CO].



\bibitem{Nojiri:2019fft}
S.~Nojiri, S.~D.~Odintsov and V.~K.~Oikonomou,
Phys. Dark Univ. \textbf{29} (2020), 100602
doi:10.1016/j.dark.2020.100602 [arXiv:1912.13128 [gr-qc]].



\bibitem{Delubac:2014aqe}
T.~Delubac \textit{et al.} [BOSS],
Astron. Astrophys. \textbf{574} (2015), A59
doi:10.1051/0004-6361/201423969 [arXiv:1404.1801 [astro-ph.CO]].




\bibitem{Hwang:2005hb}
  J.~c.~Hwang and H.~Noh,
  Phys.\ Rev.\ D {\bf 71} (2005) 063536
  doi:10.1103/PhysRevD.71.063536
  [gr-qc/0412126].




\bibitem{Nojiri:2006je}
  S.~Nojiri, S.~D.~Odintsov and M.~Sami,
  Phys.\ Rev.\ D {\bf 74} (2006) 046004
  doi:10.1103/PhysRevD.74.046004
  [hep-th/0605039].




\bibitem{Cognola:2006sp}
  G.~Cognola, E.~Elizalde, S.~Nojiri, S.~Odintsov and S.~Zerbini,
  Phys.\ Rev.\ D {\bf 75} (2007) 086002
  doi:10.1103/PhysRevD.75.086002
  [hep-th/0611198].



\bibitem{Nojiri:2005vv}
  S.~Nojiri, S.~D.~Odintsov and M.~Sasaki,
  Phys.\ Rev.\ D {\bf 71} (2005) 123509
  doi:10.1103/PhysRevD.71.123509
  [hep-th/0504052].


\bibitem{Nojiri:2005jg}
  S.~Nojiri and S.~D.~Odintsov,
  Phys.\ Lett.\ B {\bf 631} (2005) 1
  doi:10.1016/j.physletb.2005.10.010
  [hep-th/0508049].







\bibitem{Satoh:2007gn}
  M.~Satoh, S.~Kanno and J.~Soda,
  Phys.\ Rev.\ D {\bf 77} (2008) 023526
  doi:10.1103/PhysRevD.77.023526
  [arXiv:0706.3585 [astro-ph]].



\bibitem{Bamba:2014zoa}
  K.~Bamba, A.~N.~Makarenko, A.~N.~Myagky and S.~D.~Odintsov,
  JCAP {\bf 1504} (2015) 001
  doi:10.1088/1475-7516/2015/04/001
  [arXiv:1411.3852 [hep-th]].


\bibitem{Yi:2018gse}
  Z.~Yi, Y.~Gong and M.~Sabir,
  Phys.\ Rev.\ D {\bf 98} (2018) no.8,  083521
  doi:10.1103/PhysRevD.98.083521
  [arXiv:1804.09116 [gr-qc]].


\bibitem{Guo:2009uk}
  Z.~K.~Guo and D.~J.~Schwarz,
  Phys.\ Rev.\ D {\bf 80} (2009) 063523
  doi:10.1103/PhysRevD.80.063523
  [arXiv:0907.0427 [hep-th]].


\bibitem{Guo:2010jr}
  Z.~K.~Guo and D.~J.~Schwarz,
  Phys.\ Rev.\ D {\bf 81} (2010) 123520
  doi:10.1103/PhysRevD.81.123520
  [arXiv:1001.1897 [hep-th]].


\bibitem{Jiang:2013gza}
  P.~X.~Jiang, J.~W.~Hu and Z.~K.~Guo,
  Phys.\ Rev.\ D {\bf 88} (2013) 123508
  doi:10.1103/PhysRevD.88.123508
  [arXiv:1310.5579 [hep-th]].



\bibitem{Kanti:2015pda}
  P.~Kanti, R.~Gannouji and N.~Dadhich,
  Phys.\ Rev.\ D {\bf 92} (2015) no.4,  041302
  doi:10.1103/PhysRevD.92.041302
  [arXiv:1503.01579 [hep-th]].


\bibitem{vandeBruck:2017voa}
  C.~van de Bruck, K.~Dimopoulos, C.~Longden and C.~Owen,
  arXiv:1707.06839 [astro-ph.CO].



\bibitem{Kanti:1998jd}
  P.~Kanti, J.~Rizos and K.~Tamvakis,
  Phys.\ Rev.\ D {\bf 59} (1999) 083512
  doi:10.1103/PhysRevD.59.083512
  [gr-qc/9806085].




\bibitem{Pozdeeva:2020apf}
  E.~O.~Pozdeeva, M.~R.~Gangopadhyay, M.~Sami, A.~V.~Toporensky and S.~Y.~Vernov,
  arXiv:2006.08027 [gr-qc].

\bibitem{Fomin:2020hfh}
  I.~Fomin,
  arXiv:2004.08065 [gr-qc].

\bibitem{DeLaurentis:2015fea}
  M.~De Laurentis, M.~Paolella and S.~Capozziello,
  Phys.\ Rev.\ D {\bf 91} (2015) no.8,  083531
  doi:10.1103/PhysRevD.91.083531
  [arXiv:1503.04659 [gr-qc]].


\bibitem{Chervon:2019sey}
  S.~Chervon, I.~Fomin, V.~Yurov and A.~Yurov,
  doi:10.1142/11405



\bibitem{Nozari:2017rta}
  K.~Nozari and N.~Rashidi,
  Phys.\ Rev.\ D {\bf 95} (2017) no.12,  123518
  doi:10.1103/PhysRevD.95.123518
  [arXiv:1705.02617 [astro-ph.CO]].




\bibitem{Odintsov:2018zhw}
  S.~D.~Odintsov and V.~K.~Oikonomou,
  Phys.\ Rev.\ D {\bf 98} (2018) no.4,  044039
  doi:10.1103/PhysRevD.98.044039
  [arXiv:1808.05045 [gr-qc]].


  \bibitem{Kawai:1998ab}
  S.~Kawai, M.~a.~Sakagami and J.~Soda,
  Phys.\ Lett.\ B {\bf 437}, 284 (1998)
  doi:10.1016/S0370-2693(98)00925-3
  [gr-qc/9802033].


\bibitem{Yi:2018dhl}
  Z.~Yi and Y.~Gong,
  Universe {\bf 5} (2019) no.9,  200
  doi:10.3390/universe5090200
  [arXiv:1811.01625 [gr-qc]].


\bibitem{vandeBruck:2016xvt}
  C.~van de Bruck, K.~Dimopoulos and C.~Longden,
  Phys.\ Rev.\ D {\bf 94} (2016) no.2,  023506
  doi:10.1103/PhysRevD.94.023506
  [arXiv:1605.06350 [astro-ph.CO]].


\bibitem{Kleihaus:2019rbg}
  B.~Kleihaus, J.~Kunz and P.~Kanti,
  arXiv:1910.02121 [gr-qc].





\bibitem{Bakopoulos:2019tvc}
  A.~Bakopoulos, P.~Kanti and N.~Pappas,
  Phys.\ Rev.\ D {\bf 101} (2020) no.4,  044026
  doi:10.1103/PhysRevD.101.044026
  [arXiv:1910.14637 [hep-th]].


\bibitem{Maeda:2011zn}
  K.~i.~Maeda, N.~Ohta and R.~Wakebe,
  Eur.\ Phys.\ J.\ C {\bf 72} (2012) 1949
  doi:10.1140/epjc/s10052-012-1949-6
  [arXiv:1111.3251 [hep-th]].






\bibitem{Bakopoulos:2020dfg}
  A.~Bakopoulos, P.~Kanti and N.~Pappas,
  arXiv:2003.02473 [hep-th].


\bibitem{Ai:2020peo}
W.~Ai,
[arXiv:2004.02858 [gr-qc]].



\bibitem{Odintsov:2019clh}
  S.~D.~Odintsov and V.~K.~Oikonomou,
  Phys.\ Lett.\ B {\bf 797} (2019) 134874
  doi:10.1016/j.physletb.2019.134874
  [arXiv:1908.07555 [gr-qc]].



\bibitem{Oikonomou:2020oil}
V.~K.~Oikonomou and F.~P.~Fronimos,
[arXiv:2007.11915 [gr-qc]].

\bibitem{Odintsov:2020xji}
S.~D.~Odintsov, V.~K.~Oikonomou and F.~P.~Fronimos,
Annals Phys. \textbf{420} (2020), 168250
doi:10.1016/j.aop.2020.168250 [arXiv:2007.02309 [gr-qc]].



\bibitem{Oikonomou:2020sij}
V.~K.~Oikonomou and F.~P.~Fronimos,
[arXiv:2006.05512 [gr-qc]].



\bibitem{Odintsov:2020zkl}
S.~D.~Odintsov and V.~K.~Oikonomou,
Phys. Lett. B \textbf{805} (2020), 135437
doi:10.1016/j.physletb.2020.135437 [arXiv:2004.00479 [gr-qc]].


\bibitem{Odintsov:2020sqy}
S.~D.~Odintsov, V.~K.~Oikonomou and F.~P.~Fronimos,
[arXiv:2003.13724 [gr-qc]].




\bibitem{Odintsov:2020mkz}
S.~D.~Odintsov, V.~K.~Oikonomou, F.~P.~Fronimos and
S.~A.~Venikoudis,
Phys. Dark Univ. \textbf{30} (2020), 100718
doi:10.1016/j.dark.2020.100718 [arXiv:2009.06113 [gr-qc]].


\bibitem{Easther:1996yd}
  R.~Easther and K.~i.~Maeda,
  Phys.\ Rev.\ D {\bf 54} (1996) 7252
  doi:10.1103/PhysRevD.54.7252
  [hep-th/9605173].

\bibitem{Antoniadis:1993jc}
  I.~Antoniadis, J.~Rizos and K.~Tamvakis,
  Nucl.\ Phys.\ B {\bf 415} (1994) 497
  doi:10.1016/0550-3213(94)90120-1
  [hep-th/9305025].

\bibitem{Antoniadis:1990uu}
I.~Antoniadis, C.~Bachas, J.~R.~Ellis and D.~V.~Nanopoulos,
Phys.\ Lett.\ B \textbf{257} (1991), 278-284
doi:10.1016/0370-2693(91)91893-Z




\bibitem{Kanti:1995vq}
P.~Kanti, N.~Mavromatos, J.~Rizos, K.~Tamvakis and E.~Winstanley,
Phys. Rev. D \textbf{54} (1996), 5049-5058
doi:10.1103/PhysRevD.54.5049 [arXiv:hep-th/9511071 [hep-th]].



\bibitem{Kanti:1997br}
P.~Kanti, N.~Mavromatos, J.~Rizos, K.~Tamvakis and E.~Winstanley,
Phys. Rev. D \textbf{57} (1998), 6255-6264
doi:10.1103/PhysRevD.57.6255 [arXiv:hep-th/9703192 [hep-th]].




\bibitem{Odintsov:2020ilr}
S.~D.~Odintsov, V.~K.~Oikonomou and F.~P.~Fronimos,
Annals Phys. \textbf{424} (2021), 168359
doi:10.1016/j.aop.2020.168359 [arXiv:2011.08680 [gr-qc]].





\bibitem{Appleby:2009uf}
S.~A.~Appleby, R.~A.~Battye and A.~A.~Starobinsky,
JCAP \textbf{06} (2010), 005 doi:10.1088/1475-7516/2010/06/005
[arXiv:0909.1737 [astro-ph.CO]].




\bibitem{Odintsov:2020thl}
S.~D.~Odintsov and V.~K.~Oikonomou,
Phys. Lett. B \textbf{807} (2020), 135576
doi:10.1016/j.physletb.2020.135576 [arXiv:2005.12804 [gr-qc]].



\bibitem{workinprogress} 
V.~K.~Oikonomou,
[arXiv:2012.01312 [gr-qc]].



\bibitem{Hu:2007nk}
  W.~Hu and I.~Sawicki,
  Phys.\ Rev.\ D {\bf 76} (2007) 064004
  [arXiv:0705.1158 [astro-ph]].


\bibitem{Bamba:2012qi}
  K.~Bamba, A.~Lopez-Revelles, R.~Myrzakulov, S.~D.~Odintsov and L.~Sebastiani,
  Class.\ Quant.\ Grav.\  {\bf 30} (2013) 015008
  [arXiv:1207.1009 [gr-qc]].




\bibitem{Odintsov:2020vjb}
S.~D.~Odintsov, V.~K.~Oikonomou, F.~P.~Fronimos and
K.~V.~Fasoulakos,
Phys. Rev. D \textbf{102} (2020) no.10, 104042
doi:10.1103/PhysRevD.102.104042 [arXiv:2010.13580 [gr-qc]].




\bibitem{Odintsov:2020qyw}
S.~D.~Odintsov, V.~K.~Oikonomou and F.~P.~Fronimos,
Phys. Dark Univ. \textbf{29} (2020), 100563
doi:10.1016/j.dark.2020.100563 [arXiv:2004.08884 [gr-qc]].








\bibitem{Moresco:2016mzx}
  M.~Moresco {\it et al.},
  JCAP {\bf 1605} (2016) 014
  doi:10.1088/1475-7516/2016/05/014
  [arXiv:1601.01701 [astro-ph.CO]].


\bibitem{Guo:2015gpa}
  R.~Y.~Guo and X.~Zhang,
  Eur.\ Phys.\ J.\ C {\bf 76} (2016) no.3,  163
  doi:10.1140/epjc/s10052-016-4016-x
  [arXiv:1512.07703 [astro-ph.CO]].


\bibitem{Stern:2009ep}
  D.~Stern, R.~Jimenez, L.~Verde, M.~Kamionkowski and S.~A.~Stanford,
  JCAP {\bf 1002} (2010) 008
  doi:10.1088/1475-7516/2010/02/008
  [arXiv:0907.3149 [astro-ph.CO]].


\bibitem{Chuang:2012qt}
  C.~H.~Chuang and Y.~Wang,
  Mon.\ Not.\ Roy.\ Astron.\ Soc.\  {\bf 435} (2013) 255
  doi:10.1093/mnras/stt1290
  [arXiv:1209.0210 [astro-ph.CO]].




\bibitem{Ahmed:2002mj}
  M.~Ahmed, S.~Dodelson, P.~B.~Greene and R.~Sorkin,
  Phys.\ Rev.\ D {\bf 69} (2004) 103523
  doi:10.1103/PhysRevD.69.103523
  [astro-ph/0209274].


\bibitem{Cardenas:2014jya}
  V.~H.~Cardenas,
  Phys.\ Lett.\ B {\bf 750} (2015) 128
  doi:10.1016/j.physletb.2015.08.064
  [arXiv:1405.5116 [astro-ph.CO]].










\end{thebibliography}
\end{document}